\documentclass[aps,prl,amsmath,amssymb,superscriptaddress,twocolumn]{revtex4-1}
\usepackage{epsfig,epstopdf} 
\usepackage{graphicx}
\usepackage{dcolumn}
\usepackage{bm}
\usepackage{bbm}
\usepackage{ulem}
\usepackage{xcolor}
\usepackage{amsmath}
\usepackage{esint}
\usepackage{slashed}
\usepackage{mathrsfs}
\usepackage{subfigure}
\usepackage{verbatim}
\usepackage{dsfont}
\usepackage{float}
\usepackage{wasysym}
\usepackage{bbold}
\usepackage[breaklinks]{hyperref}
\hypersetup{colorlinks=true, linkcolor=blue, citecolor=blue, filecolor=blue, urlcolor=blue}

\newcommand{\nc}{\newcommand}
\nc{\be}{\begin{equation}} \nc{\ee}{\end{equation}}
\nc{\bea}{\begin{eqnarray}} \nc{\eea}{\end{eqnarray}}
\nc{\bean}{\begin{eqnarray*}} \nc{\eean}{\end{eqnarray*}}
\nc{\dg}{\dagger}
\nc{\ua}{\uparrow} \nc{\da}{\downarrow}
\nc{\lag}{\langle} \nc{\rag}{\rangle}

\begin{document}

\title{Tunable Fragile Topology in Floquet Systems}
\author{Rui-Xing Zhang}
\email{ruixing@umd.edu}
\affiliation{Condensed Matter Theory Center, Department of Physics, University of Maryland, College Park, Maryland 20742-4111, USA}
\affiliation{Joint Quantum Institute, University of Maryland, College Park, MD 20742, USA}
\author{Zhi-Cheng Yang}
\email{zcyang@umd.edu}
\affiliation{Joint Quantum Institute, University of Maryland, College Park, MD 20742, USA}
\affiliation{Joint Center for Quantum Information and Computer Science, University of Maryland, College Park, MD 20742, USA}

\begin{abstract}
	We extend the notion of fragile topology to periodically-driven systems. We demonstrate driving-induced fragile topology in two different models, namely, the Floquet honeycomb model and the Floquet $\pi$-flux square-lattice model. In both cases, we discover a rich phase diagram that includes Floquet fragile topological phases protected by crystalline rotation or mirror symmetries, Floquet Chern insulators, and trivial atomic phases, with distinct boundary features. Remarkably, the transitions between different phases can be feasibly achieved by simply tuning the driving amplitudes, which is a unique feature of driving-enabled topological phenomena. Moreover, corner-localized fractional charges are identified as a ``smoking-gun'' signal of fragile topology in our systems. Our work paves the way for studying and realizing fragile topology in Floquet systems.
\end{abstract}

\maketitle

{\it Introduction.}- Topological band insulators usually differ from trivial insulators by the existence of anomalous gapless boundary modes, a manifestation of their nontrivial bulk topology \cite{hasan2010colloquium,qi2011topological}. Various momentum-space topological invariants have been proposed to diagnose possible topology for a given set of energy bands, e.g. the Chern number \cite{thouless1982} and the $\mathbb{Z}_2$ index \cite{kane2005z2} etc. It was recently realized that band topology follows from a unified real-space definition in terms of an obstruction towards describing topological bands using exponentially localized and symmetric Wannier functions~\cite{soluyanov2011wannier,bradlyn2017topo}.
In contrast, trivial/atomic insulators are always capable of being ``Wannierized". This definition has led to the concept of fragile topology, where a set of bands are Wannier obstructed by themselves, yet the obstruction can be removed upon coupling to certain additional trivial bands~\cite{po2018fragile,bradlyn2019disconnected,hwang2019fragile,bouhon2019wilson,song2019fragile,
liu2019shift,song2020twisted}.
Practically, fragile topological bands are proposed to exist in magic-angle twisted bilayer graphene \cite{po2019faithful,ahn2019failure,song2019all}, and are engineered in photonic\cite{dePaz2019engineering} and phononic systems \cite{peri2020exp}.

Apart from static systems, topological phenomena also exist in systems far from equilibrium~\cite{cayssol2013floquet,oka2019floquet}. A prototypical example is periodically driven systems that are described by Floquet theory.
Floquet engineering can enable nontrivial band topology in statically trivial systems \cite{lindner2011floquet,jiang2011majorana,rechtsman2013photonic,usaj2014irradiated} and even achieve exotic topological phases without any static counterparts~\cite{kitagawa2010topo,rudner2013anomalous,titum2016anomalous}.
Existing studies, however, have only considered Floquet systems with stable Wannier obstructions \cite{nakagawa2020wannier}. It is then natural to ask whether fragile topology exists in an out-of-equilibrium setting, and whether Floquet engineering can realize such phases in statically trivial systems.

In this work, we provide an affirmative answer to the above questions by constructing two explicit lattice models, the Floquet honeycomb model and the Floquet $\pi$-flux model, to demonstrate driving-enabled Floquet fragile topology. By tuning the driving amplitudes, both models realize a variety of phases including Floquet fragile topological phases protected by crystalline rotation or mirror symmetries, Floquet Chern insulators, and trivial atomic phases. We characterize the fragile topological phases via atomic decompositions as well as the Wilson loop technique. 
As opposed to other phases in our models, all fragile phases carry additional higher-order topology with fractional corner charges, which serves as a unique and unambiguous boundary feature for fragile topology.

\begin{figure}[t]
	\includegraphics[width=0.45\textwidth]{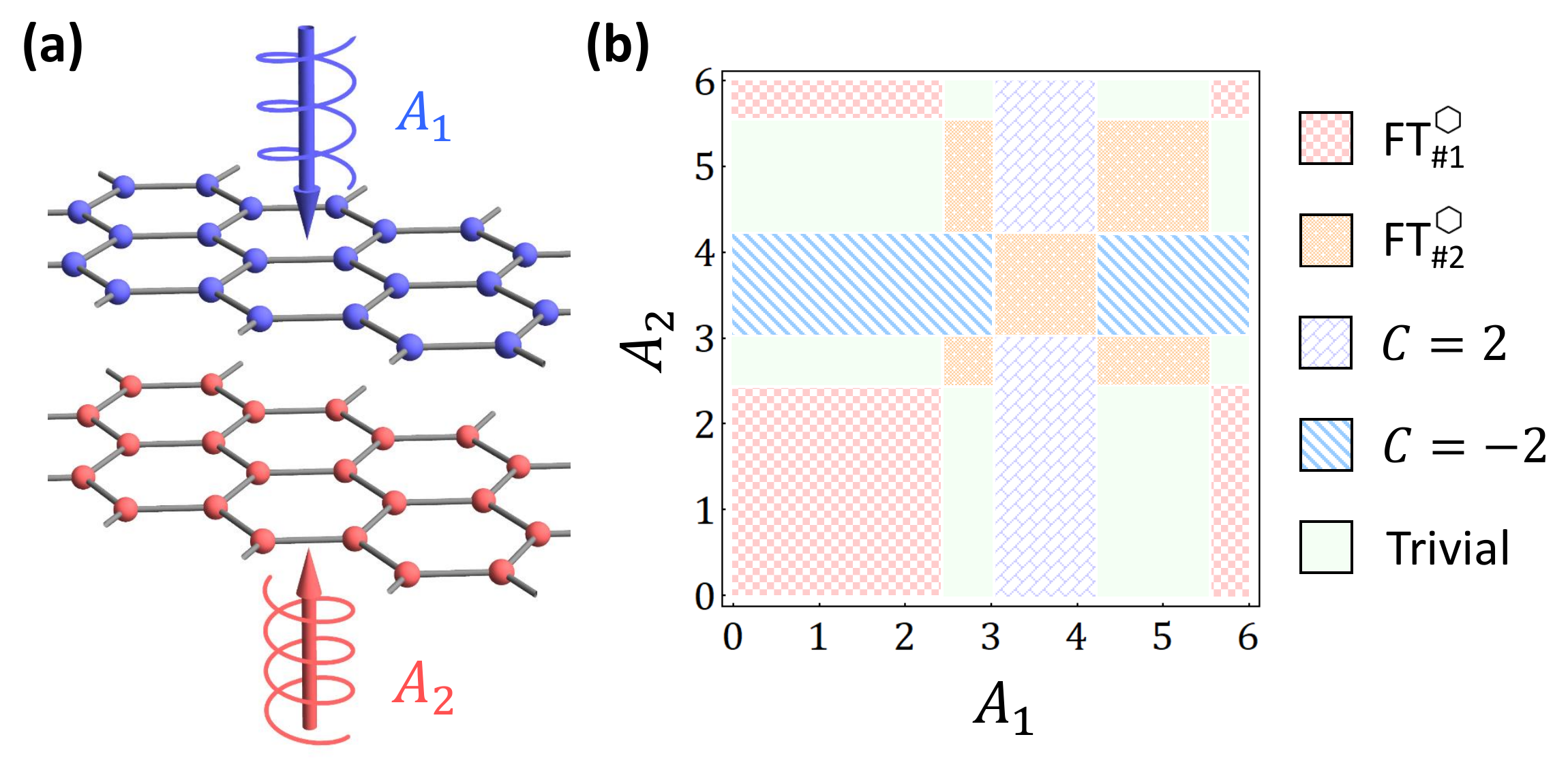}
	\caption{(a) A schematic of the Floquet honeycomb model. ``Atoms" with different colors are driven by gauge fields with different chiralities and amplitudes. (b) Phase diagram of the Floquet honeycomb model upon tuning the driving amplitudes $A_1$ and $A_2$ \cite{note1}.}
	\label{Fig1}
\end{figure} 

\begin{figure*}[t]
	\includegraphics[width=0.93\textwidth]{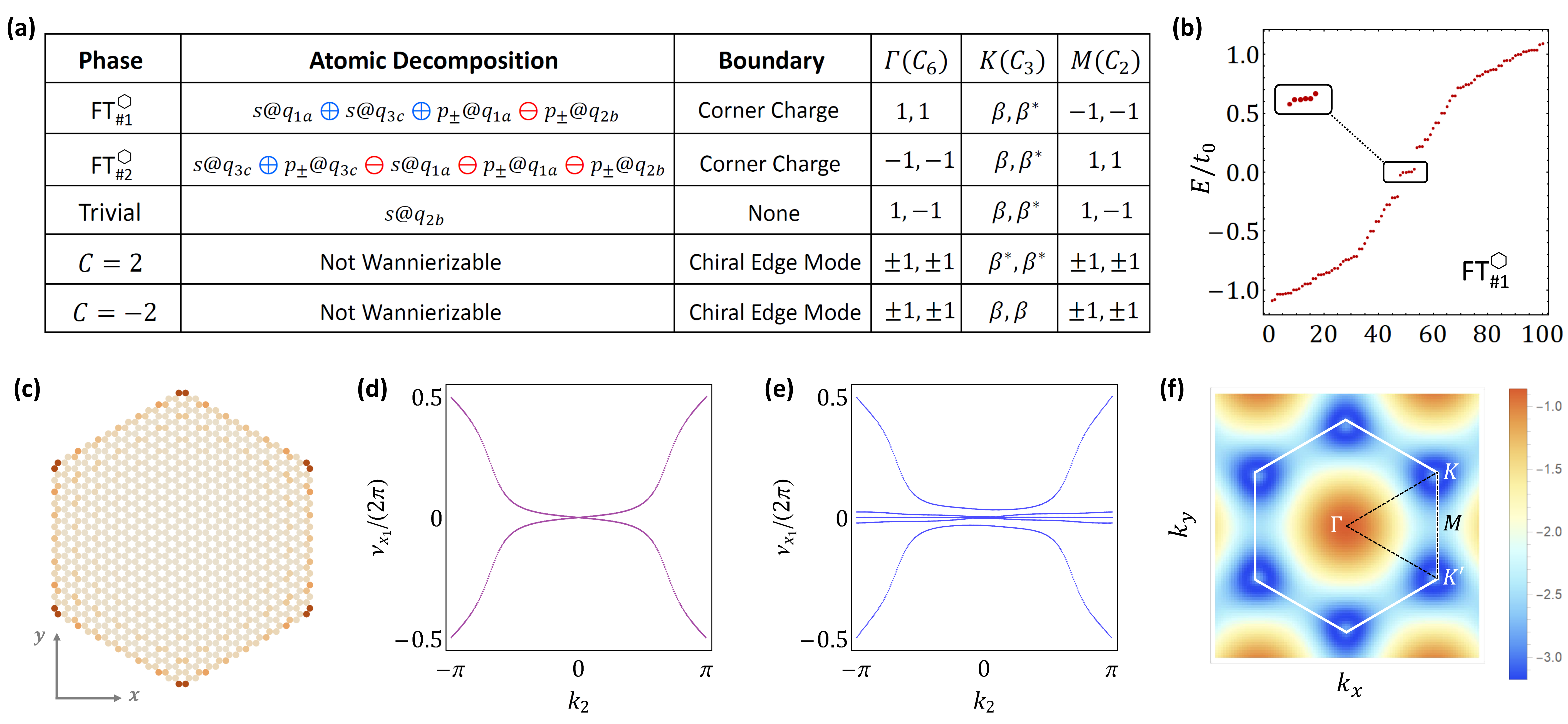}
	\caption{(a) Atomic decompositions, boundary signatures, and symmetry data for topologically distinct phases in the Floquet honeycomb model. Here $\beta=e^{i\frac{2\pi}{3}}$. (b) and (c) Fractional corner charge modes for FT$_{\#1}^{\varhexagon}$ \cite{fnote2}. (d) Windings of the Wilson loop spectrum of FT$_{\#1}^{\varhexagon}$. $k_2$ is the crystal momentum along the reciprocal lattice vector. (e) Coupling to $p_{\pm}@q_{2b}$ unwinds the Wilson loop for FT$_{\#1}^{\varhexagon}$. (f) Finite log(det$[{\cal S}({\bf k})]$) throughout the BZ implies no Wannier obstruction for FT$_{\#1}^{\varhexagon}\oplus (p_{\pm}@q_{2b})$.}
	\label{Fig2}
\end{figure*}   

{\it Floquet honeycomb model.}- Consider a two-dimensional honeycomb lattice with two species of spinless $s$-orbital electrons per site:
\begin{equation}
	H_{\varhexagon}(\tau)= -\sum_{\langle {i,j}\rangle,\alpha} t_{\alpha}(\tau) c^{\dagger}_{i, \alpha} c_{j,\alpha} + \Delta \sum_{i} c^{\dagger}_{i,1} c_{i,2} + {\rm H. c.},
\label{eq:honeycomb}
\end{equation}
where $\alpha=1, 2$ labels the species, and the hoppings $t_\alpha(\tau)$ are time-dependent.
The time-dependence in $t_\alpha$ originates from coupling to a time-periodic gauge field ${\cal A}_{\alpha}(\tau) = A_{\alpha}(\cos\omega \tau, (-1)^{\alpha}\sin \omega \tau)$ via the Peierls substitution: $ t_{\alpha}(\tau) = t_0 \text{exp}[-i\int_{{\bf r}_i}^{{\bf r}_j} {\cal A}_{\alpha}(\tau) \cdot d {\bf r}]$. Here $A_{\alpha}$ and $\omega$ are the amplitudes and frequency of the drive. This dynamic gauge field is physically equivalent to that of circularly polarized light [e.g. left-handed for $\alpha=1$ and right-handed for $\alpha=2$, as illustrated in Fig. \ref{Fig1}(a)], and can be engineered in various experimental setups~\cite{lindner2011floquet,rechtsman2013photonic,Jotzu2014}.

While $H_{\varhexagon}(\tau)$ generally breaks all its crystalline symmetries at a fixed time $\tau$, its corresponding effective Floquet Hamiltonian $H_{\varhexagon}^{F}$ generating stroboscopic time-evolutions actually restores the six-fold rotation symmetry $C_6$ in the high-frequency limit $\omega \gg |t_0|$ \cite{supplementary}, where 
\begin{equation}
	H_{\varhexagon}^{F} = h_{\widetilde{t}} (A_{\alpha}) + h_{\Delta} + h_{\lambda} (A_{\alpha},\omega) + {\cal O}(\omega^{-2}).
\end{equation}
Here $h_{\widetilde{t}}$ and $h_{\Delta}$ take the same form as in Eq.~\ref{eq:honeycomb}, now with the renormalized nearest-neighbor (NN) hopping $\widetilde{t}_{\alpha} = t_0 {\cal J}_0 (A_\alpha)$, where ${\cal J}_n(x)$ is the $n$-th Bessel function of the first kind.
$h_\lambda$ is a driving-induced Haldane-like next-nearest-neighbor (NNN) hopping term that explicitly breaks time-reversal symmetry and opens up a bulk energy gap~\cite{haldane1988model}: $h_{\lambda} (A_{\alpha},\omega)=i\sum_{\alpha}\sum_{\langle \langle i,j\rangle \rangle} \mu_{ij} \lambda_{\alpha} c^{\dagger}_{i, \alpha} c_{j,\alpha}.$
If an electron hops to its NNN (counter-)clockwise around the hexagon center, it picks up a phase $\mu_{ij} = +1 (-1)$. The NNN hopping can be analytically obtained as $\lambda_{\alpha} = -2t_0^2 \sum_{n=1}^{\infty} \sin\frac{2\pi n}{3} {\cal J}_n^2(A_{\alpha})/(n \omega)$ \cite{oka2019floquet}. Crucially, since $\widetilde{t}_{\alpha}$ and $\lambda_{\alpha}$ depend on the driving amplitudes $A_\alpha$ through the oscillating Bessel functions, their signs
can be {\it individually} controlled by tuning $A_{\alpha}$, which is the key for realizing various topological phases shown in Fig. \ref{Fig1}(b). In particular, when $\lambda_1\lambda_2<0$, $H_{\varhexagon}^F$ achieves two inequivalent fragile topological phases (dubbed FT$_{\#1}^{\varhexagon}$ and FT$_{\#2}^{\varhexagon}$) with Wannier obstructions that can nevertheless be removed upon coupling to certain atomic bands. 

To understand the nature of the Wannier obstruction, we take FT$_{\#1}^{\varhexagon}$ with $\widetilde{t}_{1,2}>0$ as an example. A similar analysis can be performed for the FT$_{\#2}^{\varhexagon}$ with $\widetilde{t}_{1,2}<0$. The first evidence of the ``fragile" Wannier obstruction manifests in the symmetry eigenvalues of the two occupied bands at high-symmetry momenta, as shown in Fig. \ref{Fig1}(a). In particular, we find that no two-band atomic insulator on a honeycomb lattice shares the same set of symmetry data as FT$_{\#1}^{\varhexagon}$~\cite{supplementary}. This clearly demonstrates an obstruction towards adiabatically connecting FT$_{\#1}^{\varhexagon}$ to an atomic system with the same amount of degrees of freedom.

We now show that such Wannier obstruction in FT$_{\#1}^{\varhexagon}$ can be removed upon adding additional atomic orbitals.
Consider coupling FT$_{\#1}^{\varhexagon}$ with a pair of $p_{\pm}$ orbitals placed at the hexagon corners (i.e. maximal Wyckoff position $q_{2b}$), which we denote as ``$p_{\pm}@ q_{2b}$" for short. Here $p_{\pm} = p_x \pm i p_y$. This composite system FT$_{\#1}^{\varhexagon}\oplus (p_{\pm}@q_{2b})$ shares the same symmetry data as the following combination of atomic insulators: $(s@q_{1a}) \oplus (s@q_{3c}) \oplus (p_{\pm}@q_{1a})$, where $q_{1a}$ and $q_{3c}$ denote the maximal Wyckoff positions at the hexagon center and the center of each hexagon edge, respectively. This implies that the composite system FT$_{\#1}^{\varhexagon}\oplus (p_{\pm}@ q_{2b})$ can be Wannierized with no obstruction. We further study the Wilson loop spectrum of FT$_{\#1}^{\varhexagon}$, as shown in Fig. \ref{Fig2}(e). The Wilson loop spectrum is gapless and features nontrivial windings, a direct implication of Wannier obstruction \cite{bouhon2019wilson,alexandradinata2014wilson}. On the other hand, when coupled to additional orbitals $p_{\pm}@ q_{2b}$, the Wilson loop for the composite system indeed unwinds and becomes gapped, as shown in Fig. \ref{Fig2}(f). This again signals the Wannierizability of the composite system FT$_{\#1}^{\varhexagon}\oplus (p_{\pm}@q_{2b})$.

Finally, we construct a set of localized trial Wannier basis $|w_l({\bf r})\rangle$ ($l=1,...,6$) for $(s@q_{1a}) \oplus (s@q_{3c}) \oplus (p_{\pm}@q_{1a})$, by superposing the tight-binding basis of $H_{\varhexagon}^F$. We provide the explicit expressions of $|w_l({\bf k})\rangle$, the Fourier transform of $|w_l({\bf r})\rangle$, in \cite{supplementary}. We now follow the procedure in Ref.~\cite{soluyanov2011wannier} and define an overlap matrix ${\cal S}_{ll'}({\bf k}) = \langle \Phi_l({\bf k})|\Phi_{l'}({\bf k})\rangle$, where $|\Phi_l({\bf k})\rangle$ is constructed by projecting our trial basis $|w_l({\bf k})\rangle$ onto the occupied states of FTI$_{\#1}\oplus (p_{\pm}@q_{2b})$. In Fig.~\ref{Fig2}(g), we map out (the logarithm of) det[${\cal S}({\bf k})$] for FTI$_{\#1}\oplus (p_{\pm}@q_{2b})$ and find det$[{\cal S}({\bf k})]\neq0$ throughout the Brillouin zone (BZ), indicating no obstruction towards a Wannier representation with $|w_l({\bf k})\rangle$ for our target system. This unambiguously proves the Wannierizability of FT$_{\#1}^{\varhexagon}\oplus (p_{\pm}@q_{2b})$ and its adiabatic equivalence to $(s@q_{1a}) \oplus (s@q_{3c}) \oplus (p_{\pm}@q_{1a})$. Formally, FT$_{\#1}^{\varhexagon}$ can be decomposed into a superposition of atomic insulators as 
\begin{equation}
	\text{FT}_{\#1}^{\varhexagon} \equiv (s@q_{1a}) \oplus (s@q_{3c}) \oplus (p_{\pm}@q_{1a}) \ominus (p_{\pm}@q_{2b}),
	\label{Eq: Atomic Decomposition}
\end{equation}
where the atomic substraction ``$\ominus$" indicates the fragile topological nature of FT$_{\#1}^{\varhexagon}$. We emphasize that the fragile topology here is protected by the $C_6$ symmetry, resembling a Floquet version of the ``shift insulator" \cite{liu2019shift}. In \cite{supplementary}, we show that breaking $C_6$ down to $C_3$ will necessarily spoil the fragile topology in our system.  

When $\lambda_1\lambda_2<0$ and $\widetilde{t}_1 \widetilde{t}_2<0$, $H_{\varhexagon}^F$ is trivially atomic, equivalent to $s@q_{2b}$, with a gapped boundary. When $\lambda_1\lambda_2>0$, both electrons species carry identical Chern numbers $|{\cal C}_{1,2}|=1$, leading to a Chern insulator phase for $H_{\varhexagon}^F$ with ${\cal C}=\pm 2$. Such a Chern insulator hosts a pair of chiral edge modes circulating along the system boundary, with the chirality (i.e. the sign of ${\cal C}_{\alpha}$) determined by the sign of $\lambda_{\alpha}$.

Remarkably, both fragile phases are in fact {\it higher-order topological}, which feature in-gap fractional corner charges $\frac{e}{6}$ (mod $e$) in a finite system with $C_6$ symmetric boundary, as is numerically confirmed in Fig. \ref{Fig2}(b)\&(d). The fractional quantization of the corner charge can be understood from our atomic decomposition, along with the fact that atomic insulators with orbitals on $q_{2b}$ and $q_{3c}$ host corner charges of $\frac{2e}{3}$ and $\frac{e}{2}$, respectively~\cite{PhysRevB.99.245151}.
The presence of robust corner charge modes clearly distinguishes FT$_{\#1,2}^{\varhexagon}$ from both the Chern insulator phases and the trivial phase, and thus serves as a clear experimental indicator of fragile topology in our model. 


\begin{figure*}
\includegraphics[width=0.97\textwidth]{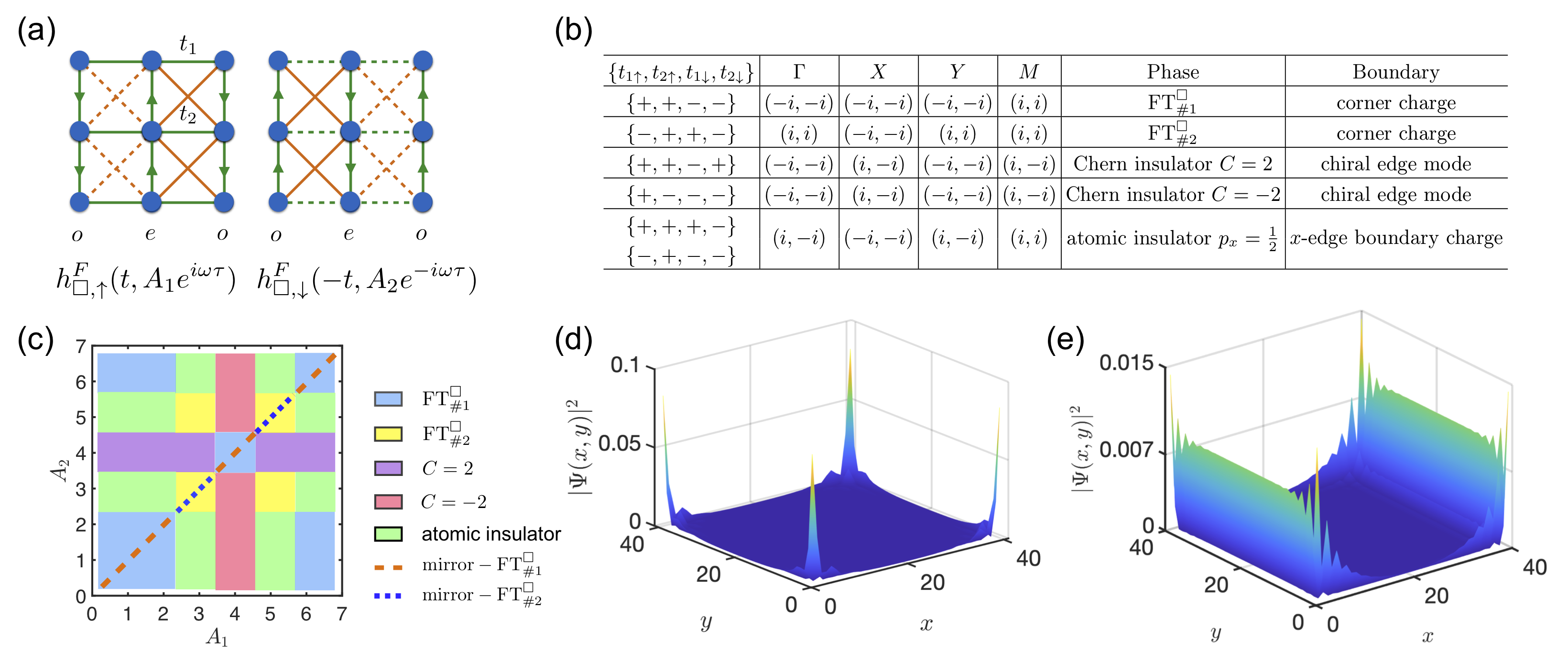}
\caption{(a) FT insulator constructed by coupling two copies of the $\pi$-flux model with opposite spins and oppositely polarized drives. Fermions hopping across the solid lines, dotted lines, and along the arrows will pick up a phase of 0, $\pi$, and $\frac{\pi}{2}$, respectively. (b) $C_2$ eigenvalues at high symmetry momenta, the corresponding phases, as well as their boundary signatures. $\{ \pm \}$ denotes the sign of each hopping term. (c) Phase diagram of Hamiltonian~(\ref{eq:pi_flux}) upon tuning $A_1$ and $A_2$. (d) Corner charge in the mirror-FT$_{\#1}^{\square}$ phase. (e) Boundary charge in the atomic insulator phase with ${\bf P}=(\frac{1}{2},0)$.}
\label{fig:pi_flux}
\end{figure*}

\textit{Floquet $\pi$-flux model.-} We now present a second example with tunable fragile topology under periodic driving.
Consider a square lattice with $\pi$-flux penetrating each elementary plaquette, with the Hamiltonian $h_{\square} = \sum_{\langle ij \rangle} t_{ij} c^\dagger_i c_j + {\rm H.c.}$. We choose the gauge such that the NN hoppings $t_{ij} = t$ on horizontal bonds, and $t_{ij} = \pm it$ on vertical bonds, as shown in Fig.~\ref{fig:pi_flux}(a). In the absence of driving, the energy spectrum of the $\pi$-flux model takes the form: $E_{\bf k} = \pm 2t \sqrt{{\rm cos}^2k_x + {\rm sin}^2 k_y}$ where ${\bf k}  \in [-\frac{\pi}{2}, \frac{\pi}{2}) \times [-\pi, \pi)$. The spectrum has two Dirac nodes at high-symmetry points $X$ and $M$ in the BZ. Similarly to the honeycomb model, the $\pi$-flux model is coupled to a time-dependent gauge field $Ae^{i\omega \tau}$ via the Peierls substitution. In the high-frequency limit, we show that the effective Floquet Hamiltonian to order $1/\omega$ is given by~\cite{supplementary}:
\begin{equation}
h_{\square}^F = \sum_{\langle ij \rangle} t_{1, ij} c^\dagger_i c_j + \sum_{\langle \langle ij \rangle \rangle} t_{2, ij} c^\dagger_i c_j + {\rm H.c.} + \mathcal{O}(\omega^{-2}),
\end{equation}
where the renormalized NN hopping and the driving-induced NNN hopping are given by $t_{1, ij} = t_{ij} \mathcal{J}_0(A)$ and $t_{2,ij} = \frac{4 \eta_{ij} t^2}{\omega} \sum_{m>0} \frac{\mathcal{J}^2_m(A)}{m} {\rm sin} \left(\frac{\pi}{2}m \right)$. The NNN hopping has alternating signs $\eta_{ij}=\pm 1$ as illustrated in Fig.~\ref{fig:pi_flux}(a). In the presence of $t_2$, each triangle in Fig.~\ref{fig:pi_flux} has a flux $\pm \frac{\pi}{2}$ which breaks time-reversal symmetry and opens up a gap at the Dirac points, yielding two bands carrying Chern numbers ${\cal C}=\pm 1$~\cite{PhysRevB.39.11413}.

Consider coupling two copies of the $\pi$-flux model with $s$-orbital electrons of \textit{opposite} spins under oppositely polarized drives:
\begin{equation}
H_{\square}^F = h_{\square, \uparrow}^F(t, A_1e^{-i \omega \tau}) + h_{\square, \downarrow}^F(-t, A_2 e^{i \omega \tau}) + H_g,
\label{eq:pi_flux}
\end{equation}
where $\pm t$ denotes the NN hoppings of the undriven model, and $H_g$ is the coupling between two copies, whose explict form will be specified below. Hamiltonian~(\ref{eq:pi_flux}) has a $C_2$ rotation symmetry with the rotation center being the bond center. 
Once again, the signs and values of NN and NNN hoppings in each copy $\{ t_{1\uparrow}, t_{2 \uparrow}, t_{1 \downarrow}, t_{2\downarrow} \}$ can be varied by tuning the two driving amplitudes $A_1$ and $A_2$, such that  Hamiltonian~(\ref{eq:pi_flux}) realizes a variety of phases with distinct boundary signatures. In Fig.~\ref{fig:pi_flux}(b)\&(c), we show the phase diagram of Hamiltonian~(\ref{fig:pi_flux}) upon tuning $A_1$ and $A_2$, as well as their boundary signatures.

For $t_{1\uparrow} t_{1\downarrow}<0$, $t_{2\uparrow} t_{2\downarrow}<0$, Hamiltonian~(\ref{eq:pi_flux}) realizes $C_2$-protected fragile topology. This further splits into two distinct phases which we label as ${\rm FT}_{\#1}^{\square}$ and ${\rm FT}_{\#2}^{\square}$, as shown in Fig.~\ref{fig:pi_flux}(b). Since $t_2$ has opposite signs in the two copies, the two occupied bands carry Chern numbers ${\cal C}=\pm 1$, and host two counter-propagating chiral edge modes. However, since the total Chern number vanishes, these chiral edge modes can be gapped out by adding $C_2$-preserving couplings between the two copies
$H_g = \sum_{ i \in \Lambda_o}  g_1 (\ c^\dagger_{i \uparrow} c_{i \downarrow} - \ c^\dagger_{i+\hat{{\bm x}} \uparrow}c_{i+\hat{{\bm x}} \downarrow}  ) +\ g_2 (\ c^\dagger_{i \uparrow} c_{i+\hat{{\bm x}} \downarrow} - \ c^\dagger_{i \downarrow} c_{i+\hat{{\bm x}} \uparrow})  + {\rm H.c.}$,
where $\Lambda_o$ denotes the sublattice belonging to odd columns, and site $i+\hat{\bm x}$ belongs to even columns. By constructing the low-energy effective edge theory~\cite{supplementary}, one can show that the edge spectrum indeed becomes gapped upon adding $H_g$. Hamiltonian~(\ref{eq:pi_flux}) thus realizes an insulator without anomalous edge modes. Nonetheless, it suffers from an obstruction towards a Wannier representation. In Fig.~\ref{fig:pi_flux}(b), we list the $C_2$ eigenvalues at high symmetry momenta for ${\rm FT}_{\#1}^{\square}$ and ${\rm FT}_{\#2}^{\square}$. We find that these symmetry representations cannot be realized with any two-band atomic insulator and hence the system is Wannier obstructed. The Wannier obstruction can also be detected from the Wilson loop spectrum. In fact, the symmetry data of ${\rm FT}_{\#1}^{\square}$ and ${\rm FT}_{\#2}^{\square}$ in Fig.~\ref{fig:pi_flux}(b) enforce nontrivial windings of the Wilson loop spectrum~\cite{PhysRevB.89.155114}, which we verify numerically~\cite{supplementary}.


Next, we show that the Wannier obstruction is fragile and can be removed upon adding atomic orbitals to Hamiltonian~(\ref{eq:pi_flux}). Consider coupling ${\rm FT}_{\#1}^{\square}$ to two spin-up orbitals per unit cell at $C_2$-symmetric positions: $(0, \frac{1}{4})$ away from the odd sublattice and $(0, -\frac{1}{4})$ away from the even sublattice. We find that the new composite system shares the same symmetry data as the atomic insulator: $(s_\uparrow @ q_{1a}) \oplus (s_\uparrow @ q_{1b}) \oplus (s_\uparrow @ q_{1c}) \oplus (s_\downarrow @ q_{1d})$, indicating that the composite system has no Wannier obstruction. Here the maximal Wyckoff positions for a $C_2$-symmetric unit cell are: $q_{1a}=(0,0), q_{1b}=(\frac{1}{2},0), q_{1c}=(0,\frac{1}{2}), q_{1d}=(\frac{1}{2},\frac{1}{2})$. Furthermore, we show in \cite{supplementary} that the Wilson loop unwinds upon coupling to the added atomic orbitals. Taken together, the Wannier obstructions in ${\rm FT}_{\#1}^{\square}$ and ${\rm FT}_{\#2}^{\square}$ are indeed fragile.

Along the line $A_1 = A_2$ in Fig.~\ref{fig:pi_flux}(c), the system has two additional mirror symmetries $M_x$ and $M_y$ \cite{ftnote3}. Under the basis $(c_{o\uparrow}^\dagger, c_{e\uparrow}^\dagger, c_{o\downarrow}^\dagger, c_{e\downarrow}^\dagger)$, these two symmetries can be written as: $M_x = -i \sigma^y \otimes \gamma^y$ and $M_y = i \sigma^z \otimes \gamma^x$ with $[M_x, M_y]=0$, where the Pauli matrices $\sigma^a$ and $\gamma^b$ act on sublattice and spin degrees of freedom respectively. The mirror symmetries protect {\it two distinct new fragile topological phases}: mirror-FT$_{\#1}^{\square}$ and mirror-FT$_{\#2}^{\square}$, whose $M_x$ and $M_y$ representations cannot be realized in any atomic insulator. We provide the atomic decompositions in the presence of mirror symmetries in~\cite{supplementary}.
Remarkably, mirror-FT$_{\#1}^{\square}$ and mirror-FT$_{\#2}^{\square}$ host four corner modes with charge $\frac{e}{4}$ (mod $e$) [Fig.~\ref{fig:pi_flux}(d)], which can be seen from the atomic decomposition and the filling anomaly~\cite{PhysRevB.99.245151}. These corner charges persist for ${\rm FT}_{\#1}^{\square}$ and ${\rm FT}_{\#2}^{\square}$ when $A_1$ and $A_2$ are slightly off the mirror-symmetric line. Nonetheless, when $M_{x,y}$ are strongly broken, $C_2$ symmetry alone will generically protect only two corner charges $\frac{e}{2}$ (mod $e$) for ${\rm FT}_{\#1}^{\square}$ and ${\rm FT}_{\#2}^{\square}$.




When $t_{2\uparrow} t_{2\downarrow}>0$, the occupied bands of $H_{\square}$ have a net Chern number ${\cal C}=\pm 2$, hence realize a Chern insulator with two chiral edge modes at the boundary. Finally, when $t_{1\uparrow} t_{1\downarrow}>0$, $t_{2\uparrow} t_{2\downarrow}<0$, $H_{\square}$ realizes an atomic insulator $(s_\uparrow @ q_{1c}) \oplus (s_\downarrow @ q_{1d})$, which has a nonzero bulk polarization ${\bf P}=(\frac{1}{2},0)$. Consequently, this atomic insulator has charge accumulation along edges parallel to the $y$-direction, but not on the other edges [Fig.~\ref{fig:pi_flux}(e)]. Therefore, similar to the honeycomb model, different phases of the $\pi$-flux model can also be diagnosed solely from their boundary signatures.


{\it Discussions} - We present two examples of Floquet systems exhibiting driving-induced tunable Floquet fragile topology with characteristic boundary features. Floquet systems are advantageous in their tunability, which offers a unique opportunity to explore the topological quantum criticality for fragile topology in experiments. Although both of our models are statically semimetallic, one may as well start from static gapped systems. Meanwhile, general fragile topological phases are recently proposed to host in-gap spectral flows under twisted boundary conditions in real space \cite{song2020twisted,peri2020exp}. For Floquet fragile topological systems, we expect that the Floquet drive acts as an on-off controller of the twisted-boundary-induced spectral flows. A detailed discussion will be left for future studies.        
 
Floquet Haldane phase, the key ingredient of our honeycomb model, has already been proposed and realized in a variety of experimental settings, including photonic waveguides~\cite{rechtsman2013photonic}, cold atoms~\cite{Jotzu2014}, and acoustic crystals~\cite{fleury2016floquet}. To realize our Floquet honeycomb model, for example, one can simply stack two layers of the acoustic Floquet Haldane system proposed in Ref. \cite{fleury2016floquet} with layer-dependent drives, and further introduce an additional acoustic waveguide at each site to bridge between the layers, which mimics the interlayer coupling $\Delta$. With the state-of-the-art fabrication techniques in the metamaterial platforms, it is conceivable that fragile topology in out-of-equilibrium systems that we propose in this work can be realized experimentally in the near future.

{\it Acknowledgment.}- We thank Eslam Khalaf, Jiabin Yu and Mohammad Hafezi for helpful discussions. We thank S. Das Sarma for a critical reading of the manuscript and useful feedback.
R.-X.Z. is supported by a JQI Postdoctoral Fellowship, the Laboratory for Physical Sciences and Microsoft. Z.-C. Y. acknowledges funding by the DoE BES Materials and Chemical Sciences Research for Quantum Information Science program (award No. DE-SC0019449), DoE ASCR FAR-QC (award No. DE-SC0020312), DoE ASCR Quantum Testbed Pathfinder program (award No. DE-SC0019040), NSF PFCQC program, AFOSR MURI, AFOSR, ARO MURI, ARL CDQI, and NSF PFC at JQI. Z.-C. Y. is also supported by AFOSR MURI FA9550-19-1-0399 and ONR MURI.

\bibliography{fragile}

\begin{thebibliography}{40}
\bibitem [{not()}]{note1}%
  \BibitemOpen
  \href@noop {} {}\bibinfo {note} {The topological phase diagram is mapped in
  the weakly coupling limit $\Delta\rightarrow 0$. Including a finite $\Delta$
  will quantitatively modify the phase diagram}\BibitemShut {NoStop}%
\bibitem [{fno()}]{fnote2}%
  \BibitemOpen
  \href@noop {} {}\bibinfo {note} {The parameters for Floquet honeycomb model
  are $t_0=2, A=1.7,\omega=15$. When calculating the corner charge modes, we
  have artificially enlarged the energy gap to overcome the finite size
  effect}\BibitemShut {NoStop}%
\bibitem [{ftn()}]{ftnote3}%
  \BibitemOpen
  \href@noop {} {}\bibinfo {note} {We note that the Floquet honeycomb model
  also possesses in-plane mirror symmetries along $A_1=A_2$ that provide
  additional symmetry protections to the fragile topology. Such mirror
  symmetries can be removed by including a species-dependent on-site
  potential}\BibitemShut {NoStop}%
\end{thebibliography}


\newpage
\onecolumngrid
\appendix

\subsection*{\large Supplemental Material for ``Tunable Fragile Topology in Floquet Systems"}

\section{Appendix A: Symmetry Properties of Floquet Honeycomb Model}

\subsection{Symmetry of Static Honeycomb Model}

We start by briefly reviewing the symmetry properties of the Floquet honeycomb model in the zero driving limit $A_{1,2}\rightarrow 0$. In this case, the model Hamiltonian resembles that of a $A$-$A$ stacked bilayer graphene. In momentum space, we have 
\begin{equation}
H_{\varhexagon}^s({\bf k}) = -t_0 (\text{Re}[f({\bf k})]\mu_0\otimes \sigma_x + \text{Im}[f({\bf k})] \mu_0\otimes \sigma_y) + \Delta \mu_x\otimes \sigma_0,
\end{equation}
where the superscript ``$s$" indicates the static limit. We have defined the nearest-neighbor hopping function $f({\bf k}) = \sum_{i=1}^3 e^{-i {\bf k} \cdot \delta_i}$ with $\delta_1 = \frac{1}{2}(1,\sqrt{3})$, $\delta_2 = \frac{1}{2}(1,-\sqrt{3})$, and $\delta_3 = (-1,0)$. The Pauli matrices $\mu_i$ and $\sigma_i$ ($i=0,x,y,z$) act on the species and sublattice degrees of freedom, respectively.

We will focus on rotational symmetries and ignore the mirror symmetries of the system, since they will be explicitly broken even in the high-frequency limit. The six-fold rotation $C_6 = \mu_0 \otimes \sigma_x$ exchanges the two sublattices of the honeycomb lattice and $H^s_{\varhexagon}({\bf k})$ transforms under $C_6$ as
\begin{equation}
C_6 H_{\varhexagon}^s({\bf k}) C_6^{\dagger} = H_{\varhexagon}^s ( R_6 {\bf k}),
\end{equation}  
with
\begin{equation}
R_n = \begin{pmatrix}
\cos \frac{2\pi}{n} & -\sin  \frac{2\pi}{n} \\
\sin  \frac{2\pi}{n} & \cos  \frac{2\pi}{n} \\
\end{pmatrix}.
\end{equation}
In the Brillouin zone, $\Gamma=(0,0)$ is invariant under $C_6$. We have two inequivalent $K= \frac{2\pi}{3\sqrt{3}}(\sqrt{3},1)$ and $K'= \frac{2\pi}{3\sqrt{3}}(\sqrt{3},-1)$ that have little group $C_3$. In addition, there are three $M= \frac{2\pi}{3}(1,0)$, $M'= \frac{\pi}{\sqrt{3}}(1,\sqrt{3})$, and $M''=\frac{\pi}{\sqrt{3}}(-1,\sqrt{3})$ that are invariant under $C_2$. 
The reciprocal lattice vectors for the honeycomb lattice are given by
\begin{equation}
{\bf G}_1 = \frac{2\pi}{3}(1,\sqrt{3}),\ \ {\bf G}_2 = \frac{2\pi}{3}(1,-\sqrt{3}).
\end{equation} 
It is easy to check that $H_{\varhexagon}^s({\bf k} + n_1 {\bf G}_1 + n_2 {\bf G}_2) \neq H_{\varhexagon}^s({\bf k})$. To make the Hamiltonian invariant under a shift of reciprocal lattice vectors, we consider a unitary transformation $V({\bf k})$
\begin{equation}
\widetilde{H}^s_{\varhexagon}({\bf k}) = V({\bf k}) H^s_{\varhexagon}({\bf k}) V({\bf k})^{\dagger}.
\end{equation}
where 
\begin{equation}
V({\bf k}) = \mu_x\otimes \begin{pmatrix}
e^{-i {\bf k}\cdot t_A} & 0 \\
0 & e^{-i {\bf k}\cdot t_B}
\end{pmatrix}.
\end{equation}
Physically, this gauge choice corresponds to choosing the hexagonal center as the unit cell origin. For a given rotation symmetry $C_n$, we have
\begin{eqnarray}
\widetilde{H}^s_{\varhexagon}(R_n {\bf k}) &=& V(R_n {\bf k}) C_n V({\bf k})^{\dagger}  V({\bf k}) H^s_{\varhexagon}({\bf k}) V({\bf k})^{\dagger} V({\bf k}) C_n^{\dagger} V(R_n {\bf k})^{\dagger} \nonumber \\
&=& V(R_n {\bf k}) C_n V({\bf k})^{\dagger}  \widetilde{H}^s_{\varhexagon}({\bf k})  V({\bf k}) C_n^{\dagger} V(R_n {\bf k})^{\dagger} \nonumber \\
&=& \widetilde{C}_n \widetilde{H}^s_{\varhexagon}({\bf k}) \widetilde{C}_n^{\dagger}.
\end{eqnarray}
Therefore, the six-fold rotation operation under the new basis is
\begin{equation}
\widetilde{C}_6({\bf k}) = V(R_6 {\bf k}) C_6 V({\bf k})^{\dagger} = \mu_0\otimes \begin{pmatrix}
0 & e^{-\frac{i}{2}(3 k_x -\sqrt{3}k_y)} \\
1 & 0 \\
\end{pmatrix}.
\end{equation}
Similarly, the three-fold and two-fold rotation operations under the new basis are
\begin{eqnarray}
\widetilde{C}_3({\bf k}) &=& V(R_3 {\bf k}) C_3 V({\bf k})^{\dagger} = \mu_0\otimes \begin{pmatrix}
e^{i \sqrt{3}k_y} & 0 \\
0 & e^{-\frac{i}{2}(3k_x - \sqrt{3}k_y)} \\
\end{pmatrix}, \nonumber \\
\widetilde{C}_2({\bf k}) &=& V(R_2 {\bf k}) C_2 V({\bf k})^{\dagger} = \mu_0\otimes e^{i \sqrt{3} k_y} \sigma_x.
\end{eqnarray}
We notice that
\begin{equation}
\widetilde{C}_3 \neq \widetilde{C}_6^2,\ \ \widetilde{C}_2 \neq \widetilde{C}_6^3.
\end{equation} 

\subsection{Emergent Lattice Symmetry and High Frequency Limit} 

\begin{figure}[t]
	\includegraphics[width=0.7\textwidth]{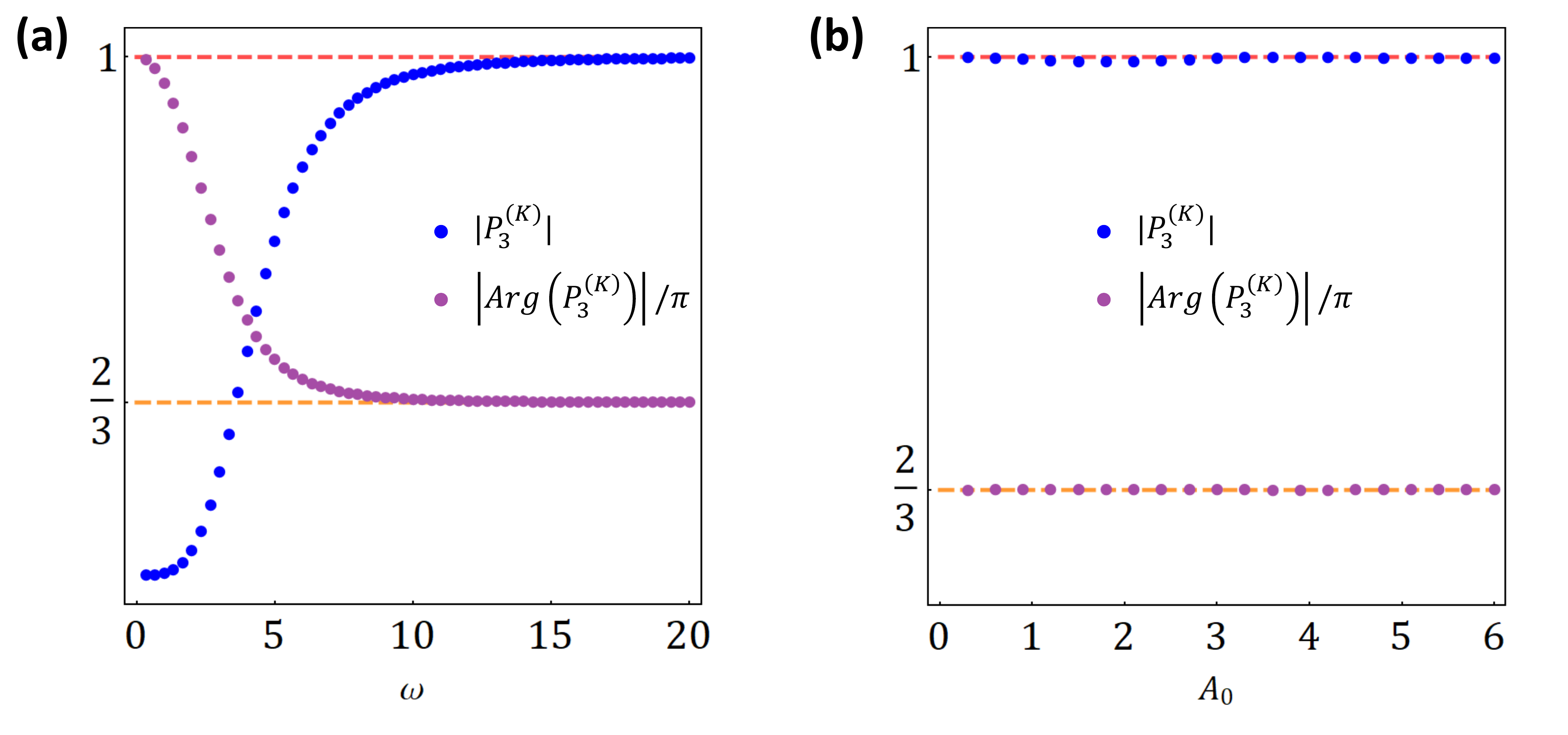}
	\caption{Effect of lattice symmetry breaking as a function of (a) $\omega$ with a fixed $A_0= 1.7$; and (b) $A_0$ with a fixed $\omega=15$.}
	\label{Fig: Floquet Symmetry Breaking}
\end{figure}  

As we have mentioned in the main text, the time-dependent Hamiltonian $H_{\varhexagon}(\tau)$ explicitly breaks all crystalline symmetries at a fixed time $\tau$. Such symmetry breaking is easy to see if we couple the dynamic gauge potential ${\bf A}(\tau) = A_0 (\cos \omega \tau, \sin \omega \tau)$ to $H_{\varhexagon}^s({\bf k})$ via a minimal coupling,
\begin{equation}
{\bf k} \rightarrow {\bf k} + {\bf A} (\tau).
\end{equation} 
Back in real space, the effective Floquet Hamiltonian for the honeycomb model is approximated via a high-frequency expansion \cite{oka2019floquet}
\begin{equation}
	H_{\varhexagon}^F = H_{\varhexagon}^{(0)} + \frac{[H_{\varhexagon}^{(1)},H_{\varhexagon}^{(-1)}]}{\omega} + {\cal O}(\frac{1}{\omega^2}),
\end{equation}
where we have defined 
\begin{eqnarray}
	H_{\varhexagon}^{(n)} = \int_0^T H_{\varhexagon}(\tau) e^{in\omega\tau} d\tau.
\end{eqnarray}
Evaluationg the above expression explicitly, we arrive at the $H_{\varhexagon}^F$ shown in the main text, which apparently has $C_6$ rotation symmetry.\\

To visualize how the rotation symmetry gradually emerges in the Floquet Hamiltonian as one increases the driving frequency, we write our Hamiltonian in frequency space, which now becomes an infinite-dimensional matrix,
\begin{equation}
	H_{\varhexagon}^F = \begin{pmatrix}
	\ddots & H_{\varhexagon}^{(-1)} & H_{\varhexagon}^{(-2)} &  & \\
	H_{\varhexagon}^{(1)} & H_{\varhexagon}^{(0)} -(m-1)\omega & H_{\varhexagon}^{(-1)} & H_{\varhexagon}^{(-2)} &  \\
	H_{\varhexagon}^{(2)} & H_{\varhexagon}^{(1)} & H_{\varhexagon}^{(0)} -m \omega & H_{\varhexagon}^{(-1)} & H_{\varhexagon}^{(-2)} &  \\
	& H_{\varhexagon}^{(2)} & H_{\varhexagon}^{(1)} & H_{\varhexagon}^{(0)} -(m+1) \omega & H_{\varhexagon}^{(-1)} &   \\
	& & H_{\varhexagon}^{(2)} & H_{\varhexagon}^{(1)} & \ddots \\
	\end{pmatrix}.
\end{equation}
For practical purpose, we truncate the frequency space Hamiltonian up to $N\omega$ with $N\gg 1$. To quantify the rotation symmetry breaking, we consider $|\psi^F_K\rangle$, an energy eigenstate of $H_{\varhexagon}^F$ at high symmetry point $K$, and evaluate 
\begin{equation}
	P_3^{(K)} = \langle \psi^F_K | C_3 | \psi^F_K \rangle.
\end{equation}
In the high-frequency limit, we find that a $C_6$-symmetric Floquet Hamiltonian will have
\begin{equation}
	|P_3^{(K)}|=1,\ \frac{1}{\pi}|\text{Arg}(P_3^{(K)})|= \frac{2\pi}{3},
\end{equation}

As shown in Fig. \ref{Fig: Floquet Symmetry Breaking} (a), we evaluate both the magnitude and the phase of $P_3^{(K)}$ as a function of $\omega$. For small $\omega$, both quantities deviate from the ideal values, which signals explicit symmetry breaking of $C_6$. In particular, $P_3^{(K)}$ reaches the expected value at $\omega\sim 12$, which indicates that the system reaches the high-frequency limit. With $\omega=15$, we also plot the $P_3^{(K)}$ as a function of $A_1=A_2=A_0$, which confirms that the emergent $C_6$ symmetry remains robust as we change the driving amplitude.

\section{Appendix B: Fragile Topology in the Honeycomb model}

\subsection{Atomic Symmetry Data and Decomposition of Fragile Phase}

In this section, we provide the symmetry data for different atomic insulators on a honeycomb lattice, based on which, we will discuss possible atomic decompositions for both fragile topological phases in the Floquet honeycomb model. By placing one atomic orbital on a maximal Wyckoff position, we obtain a corresponding atomic insulator and extract its symmetry data at high symmetry momenta. For a honeycomb lattice, there exists three inequivalent maximal Wyckoff positions: (i) the hexagon center $q_{1a}$; (ii) the hexagon corner $q_{2b}$; (iii) the midpoint of an hexagon edge $q_{3c}$. For a $C_6$-symmetric spinless system, we should, in principle, consider atomic orbitals with an orbital angular momentum $l\in \{-2,-1,0,1,2,3\}$. As shown in Table. \ref{Table: Atomic Data}, we list the atomic symmetry data for $l=0$ ($s$-orbital) and $l=\pm 1$ ($p_{\pm}$-orbital) for all maximal Wyckoff positions.

\begin{table}[t]
	\includegraphics[width=0.7\textwidth]{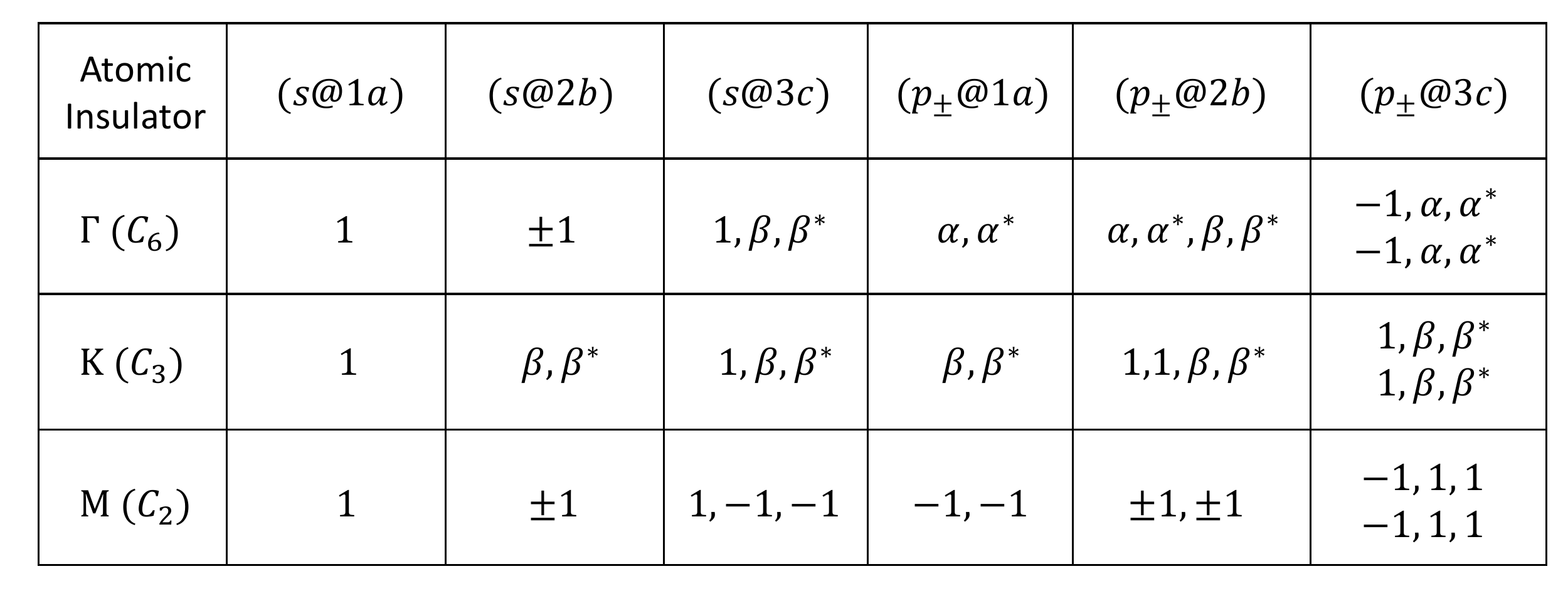}
	\caption{Symmetry data for atomic orbitals on a honeycomb lattice. $\alpha = e^{i\frac{\pi}{3}}$, $\beta = e^{i\frac{2\pi}{3}}$.}
	\label{Table: Atomic Data}
\end{table}   

To construct the atomic decomposition for the fragile topological phases, we consider a linear superposition of possible atomic insulators with integer-valued coefficients to match the symmetry data of the fragile phase. We point out that such an atomic decomposition for a given fragile phase is not necessarily unique. In other words, one can couple a fragile phase with different combinations of atomic orbitals to remove the Wannier obstruction. For example, we find two inequivalent decomposition schemes for the FT$_{\#2}^{\varhexagon}$ phase by considering {\it only} $s$-orbital and $p_{\pm}$ orbitals,
\begin{eqnarray}
\text{FT}_{\#2}^{\varhexagon} &\equiv& (s@q_{2b}) \oplus (s@q_{2b}) \oplus (p_{\pm} @q_{2b}) \ominus (s @ q_{1a}) \ominus (s @ q_{3c}) \ominus (p_{\pm} @ q_{1a})    \nonumber \\
&\equiv& (s @ q_{3c}) \oplus (p_{\pm} @q_{3c}) \ominus (s @ q_{1a}) \ominus (p_{\pm} @ q_{1a}) \ominus (p_{\pm} @ q_{2b})
\end{eqnarray}
For the first decomposition, we require adding six additional atomic orbitals $(s @ q_{1a}) \oplus (s @ q_{3c}) \oplus (p_{\pm} @ q_{1a})$ to make FT$_{\#2}^{\varhexagon}$ Wannierizable, while the second decomposition requires adding seven atomic orbitals $(s @ q_{1a}) \oplus (p_{\pm} @ q_{1a}) \oplus (p_{\pm} @ q_{2b})$ to trivialize FT$_{\#2}^{\varhexagon}$. One could in principle find more decomposition schemes if atomic orbitals with higher angular momenta are considered as well.

\subsection{Trial Wannier Basis for FT$_{\#1}^{\varhexagon}\oplus (p_{\pm}@q_{2b})$}

In this section, we provide detailed expressions for the trial Wannier basis $|w_l\rangle$ ($l=1,2,...,6$) for FT$_{\#1}^{\varhexagon}\oplus (p_{\pm}@q_{2b})$ that are used to show the absence of Wannier obstruction [e.g. Fig. 2(f) in the main text]. This choice of Wannier basis is inspired by the ones in Ref. \cite{liu2019shift}. We expect the trial Wannier basis to satisfy the following requirements:
\begin{itemize}
	\item $|w_l\rangle$ is constructed using the original tight-binding basis of FT$_{\#1}^{\varhexagon}\oplus (p_{\pm}@q_{2b})$;
	\item $|w_l\rangle$ should exactly reproduce the symmetry data of FT$_{\#1}^{\varhexagon}\oplus (p_{\pm}@q_{2b})$.
\end{itemize}
Recall that we expect 
\begin{equation}
	\text{FT}_{\#1}^{\varhexagon}\oplus (p_{\pm}@q_{2b}) \equiv (s@ q_{1a}) \oplus (s@q_{3c}) \oplus (p_{\pm}@ q_{1a}). 	
\end{equation}
Hence, the first three trial Wannier bases aim at constructing $s@q_{3c}$,
\begin{eqnarray}
	|w_1({\bf k})\rangle &=& (e^{i {\bf k \cdot t}_A}, e^{i {\bf k \cdot t}_B},e^{i {\bf k \cdot t}_A}, e^{i {\bf k \cdot t}_B},e^{i {\bf k \cdot t}_A}, -e^{i {\bf k \cdot t}_B},e^{i {\bf k \cdot t}_A}, -e^{i {\bf k \cdot t}_B})^T, \nonumber \\
	|w_2 ({\bf k})\rangle &=& C_6 |w_1 (R_6^{-1}{\bf k})\rangle,\ \	|w_3 ({\bf k})\rangle = C_3 |w_1 (R_3^{-1}{\bf k})\rangle.
\end{eqnarray}
The other three bases aim at constructing $s@q_{1a}$, $p_+ @ q_{1a}$, and $p_- @ q_{1a}$, respectively.
\begin{eqnarray}
	|w_4({\bf k})\rangle &=& (0,0,g_{0A},g_{0B},g_{-1A},g_{-1B},g_{+1A},g_{+1B})^T, \nonumber \\
	|w_5({\bf k})\rangle &=& (g_{+1A},g_{+1B},g_{+1A},g_{+1B},g_{0A},g_{0B},g_{+2A},g_{+2B})^T \nonumber \\
	|w_6({\bf k})\rangle &=& (0,0,g_{-1A},g_{-1B},g_{-2A},g_{-2B},g_{0A},g_{0B})^T.
\end{eqnarray}	
where we have defined 
\begin{eqnarray}
	g_{0A} &=& e^{i {\bf k \cdot t}_A} + e^{-i {\bf k \cdot t}_B} + e^{i {\bf k} \cdot (-{\bf t}_A + {\bf t}_B)},\ \ \ \ \ \ \ \ \ \ \ \ \ \ 
	g_{0B} = e^{i {\bf k \cdot t}_B} + e^{-i {\bf k \cdot t}_A} + e^{i {\bf k} \cdot ({\bf t}_A - {\bf t}_B)}, \nonumber \\
	g_{-1A} &=& \alpha^5 e^{i {\bf k \cdot t}_A} + \alpha^3 e^{-i {\bf k \cdot t}_B} + \alpha e^{i {\bf k} \cdot (-{\bf t}_A + {\bf t}_B)},\ \ \ \ 
	g_{-1B} = e^{i {\bf k \cdot t}_B} + \alpha^2 e^{-i {\bf k \cdot t}_A} + \alpha^4 e^{i {\bf k} \cdot ({\bf t}_A - {\bf t}_B)} \nonumber \\
	g_{+1A} &=& \alpha e^{i {\bf k \cdot t}_A} + \alpha^3 e^{-i {\bf k \cdot t}_B} + \alpha^5 e^{i {\bf k} \cdot (-{\bf t}_A + {\bf t}_B)},\ \ \ \ 
	g_{+1B} = e^{i {\bf k \cdot t}_B} + \alpha^4 e^{-i {\bf k \cdot t}_A} + \alpha^2 e^{i {\bf k} \cdot ({\bf t}_A - {\bf t}_B)} \nonumber \\
	g_{-2A} &=& \alpha^4 e^{i {\bf k \cdot t}_A} + e^{-i {\bf k \cdot t}_B} + \alpha^2 e^{i {\bf k} \cdot (-{\bf t}_A + {\bf t}_B)},\ \ \ \ \ \ 
	g_{-2B} = e^{i {\bf k \cdot t}_B} + \alpha^4 e^{-i {\bf k \cdot t}_A} + \alpha^2 e^{i {\bf k} \cdot ({\bf t}_A - {\bf t}_B)} \nonumber \\
	g_{+2A} &=& \alpha^2 e^{i {\bf k \cdot t}_A} + e^{-i {\bf k \cdot t}_B} + \alpha^4 e^{i {\bf k} \cdot (-{\bf t}_A + {\bf t}_B)},\ \ \ \ \ \ 
	g_{+2B} = e^{i {\bf k \cdot t}_B} + \alpha^2 e^{-i {\bf k \cdot t}_A} + \alpha^4 e^{i {\bf k} \cdot ({\bf t}_A - {\bf t}_B)}
\end{eqnarray}
where $\alpha = e^{i\frac{\pi}{3}}$ is a phase factor. It is easy to show that the above trial Wannier basis satisfy all the symmetry requirements.

\subsection{Composite Hamiltonian of FT$_{\#1}^{\varhexagon}\oplus (p_{\pm}@q_{2b})$ for Wilson Loop Calculation}

To show the composite system FT$_{\#1}^{\varhexagon}\oplus (p_{\pm}@q_{2b})$ does not have a nontrivial Wilson loop winding, we consider the following composite Hamiltonian,
\begin{equation}
	H_\text{composite}({\bf k}) = \begin{pmatrix}
	H_{\varhexagon}^F({\bf k}) & h_c \\
	h_c^{\dagger} &  h_{p_{\pm}@q_{2b}} \\
	\end{pmatrix}
\end{equation}
Here, the atomic system is described by
\begin{equation}
	h_{p_{\pm}@q_{2b}}({\bf k}) = -\epsilon_F\widetilde{\mu}_0\otimes\sigma_0-t_p \text{Re}[f({\bf k})]\widetilde{\mu}_0\otimes \sigma_x - t_p \text{Im}[f({\bf k})] \widetilde{\mu}_0\otimes \sigma_y,
\end{equation}
where $\widetilde{\mu}_i$ are the Pauli matrices describing $p_{\pm}$ orbital degrees of freedom. The general $4\times 4$ coupling matrix satisfying symmetry requirement is given by
\begin{equation}
	h_c ({\bf k}) = \begin{pmatrix}
	v_{11} J_{1}({\bf k}) & 0 & v_{13} J_{3}({\bf k}) & 0 \\
	0 & v_{11} J_{2}({\bf k}) & 0 & v_{13} J_{4} ({\bf k}) \\
	v_{31} J_{1}({\bf k}) & 0 & v_{33} J_{3}({\bf k}) & 0 \\
	0 & v_{31} J_{2}({\bf k}) & 0 & v_{33} J_{4} ({\bf k}) \\
	\end{pmatrix},
\end{equation}
where
\begin{eqnarray}
	J_{1}({\bf k}) &=& e^{i {\bf k\cdot b}_1} + \alpha^4 e^{i {\bf k\cdot b}_2} + \alpha^2 e^{i {\bf k\cdot b}_3} \nonumber \\
	J_{2}({\bf k}) &=& -e^{-i {\bf k\cdot b}_1} + \alpha e^{-i {\bf k\cdot b}_2} + \alpha^5 e^{-i {\bf k\cdot b}_3} \nonumber \\
	J_{3}({\bf k}) &=& e^{i {\bf k\cdot b}_1} + \alpha^2 e^{i {\bf k\cdot b}_2} + \alpha^4 e^{i {\bf k\cdot b}_3} \nonumber \\
	J_{4}({\bf k}) &=& -e^{-i {\bf k\cdot b}_1} + \alpha^5 e^{-i {\bf k\cdot b}_2} + \alpha e^{-i {\bf k\cdot b}_3}
\end{eqnarray}
Here ${\bf b}_1 = (0,\sqrt{3})$, ${\bf b}_2 =(-\frac{3}{2}, -\frac{\sqrt{3}}{2})$, and ${\bf b}_3 =(\frac{3}{2}, -\frac{\sqrt{3}}{2})$ are the displacement vectors between next-nearest neighbors atoms. When calculating the Wilson loop spectrum in Fig. 2 (d) \& (e) in the main text, we have chosen the following set of parameters
\begin{equation}
	\Delta=0, \epsilon_F=4, t_p=0.4, v_{11} = 0.8, v_{33} = 0.4, v_{13} = 0.3, v_{31}=0.5
\end{equation}

\subsection{No Fragile Topology for $C_3$-symmetric Floquet Honeycomb Model}

Now we show that when breaking the $C_6$ symmetry of $H_{\varhexagon}^F$ to $C_3$ by adding a sublattice staggered potential 
\begin{equation}
	h_\text{stagger} = \widetilde{\Delta} \mu_0\otimes \sigma_z,
\end{equation}
the original fragile topological phase will be trivialized. This directly implies that the fragile topology here is protected by $C_6$ symmetry.

Take FT$_{\#1}^{\varhexagon}$ as an example. With additional $h_\text{stagger}$ and the remaining $C_3$ symmetry, we only need to consider the $C_3$ symmetry data at $\Gamma$ and $K$. By comparing the new symmetry data of FT$_{\#1}^{\varhexagon}$ with Table. \ref{Table: Atomic Data}, we find that the new symmetry data exactly matches that of $s@q_{2b}$. To show that a $C_3$-symmetric FT$_{\#1}^{\varhexagon}$ phase is adiabatically connected to the atomic insulator $s@q_{2b}$, we consider the following trial Wannier basis $|\widetilde{w}_{1,2}\rangle$ for $s@q_{2b}$,
\begin{eqnarray}
	|\widetilde{w}_1({\bf k})\rangle &=& (e^{i {\bf k \cdot t}_A},0,-e^{i {\bf k \cdot t}_A},0)^T, \nonumber \\
	|\widetilde{w}_2({\bf k})\rangle &=& (0, e^{i {\bf k \cdot t}_B},0,e^{i {\bf k \cdot t}_B})^T.
\end{eqnarray}

\begin{figure}[t]
	\includegraphics[width=0.7\textwidth]{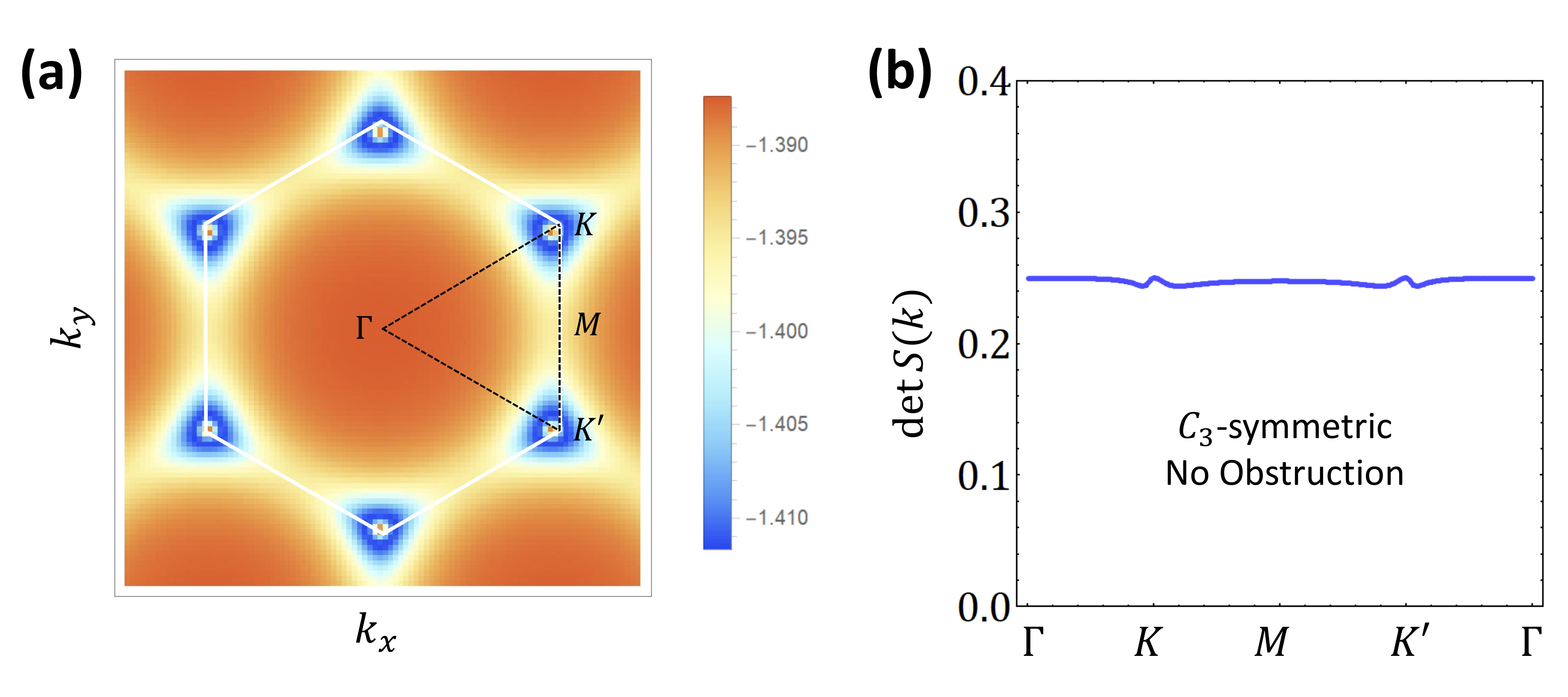}
	\caption{log(det$[{\cal S}({\bf k})]$) of FT$_{\#1}^{\varhexagon}$ in the presence of a $C_6$-breaking staggered potential (a) for the entire BZ; (b) along the high symmetry lines.}
	\label{Fig: C3}
\end{figure}

With the above trial Wannier basis, we calculate (the logarithm of) the determinant of the overlap matrix det$[{\cal S}({\bf k})]$. As shown in Fig. \ref{Fig: C3} (a) and (b), we find that the determinant remains finite throughout the BZ, which confirms the absence of an obstruction in describing $C_3$-symmetric FT$_{\#1}^{\varhexagon}$ with our trial Wannier basis. In other words, when breaking $C_6$ down to $C_3$, the FT$_{\#1}^{\varhexagon}$ phase becomes trivialized and is adiabatically connected to an atomic insulator $s@q_{2b}$.

\section{Appendix C: Floquet $\pi$-flux model}

\subsection{Effective Floquet Hamiltonian from High Frequency Expansion}
 
In this section, we derive the effective Floquet Hamiltonian~(5) in the main text for the driven $\pi$-flux model using high frequency expansion. The time-dependent Hamiltonian of the $\pi$-flux model coupled to a vector potential $\mathcal{A}(\tau) = A({\rm cos}\omega \tau, {\rm sin}\omega \tau)$ is given by:
\begin{eqnarray}
\label{eq:driven}
H(\tau) &=& \sum_{\langle i j \rangle} t_{ij}(\tau) \ c^\dagger_i c_j + {\rm H.c.}   \\
t_{ij}(\tau) &=& t_{ij} {\rm exp} \left( -i \int_{{\bm r}_i}^{{\bm r}_j}  {\mathcal A}(\tau) \cdot {\rm d} {\bm r} \right)   \nonumber  \\
&=& t_{ij} {\rm exp} \big[ -i {\mathcal A}(\tau) \cdot ({\bm r}_j - {\bm r}_i)\big].
\end{eqnarray}
The Floquet Hamiltonian $H_{F}$ is defined via the stroboscopic time evolution operator:
\begin{equation}
U_F = \mathcal{T} {\rm exp} \left( -i \int_0^T H(\tau) {\rm d} \tau \right) \equiv {\rm exp} \left( -i H_{F} T\right),
\end{equation}
where $T=\frac{2\pi}{\omega}$. To leading order in $1/\omega$, the effective Hamiltonian can be written as~\cite{bukov2015universal}:
\begin{equation}
H_{F} = H_0 + \sum_{m \neq 0} \frac{[H_{-m}, H_m]}{2m\omega} + \mathcal{O}(\omega^{-2}),
\end{equation}
where
\begin{equation}
H_0 = \frac{1}{T} \int_0^T {\rm d}\tau H(\tau),  \quad \quad H_m = \frac{1}{T} \int_0^T {\rm d}\tau H(\tau) e^{-im\omega \tau}
\end{equation}
are the Fourier components of $H(\tau)$. For Hamiltonian~(\ref{eq:driven}) with NN hopping, the only nonvanishing term in $[H_{-m}, H_m]$ gives rise to a second NN hopping. The Floquet Hamiltonian thus takes the following form: 
\begin{equation}
H_{F} = \sum_{\langle ij \rangle} t_{1, ij} c^\dagger_i c_j + \sum_{\langle \langle ij \rangle \rangle} t_{2, ij} c^\dagger_i c_j + {\rm H.c.} + \mathcal{O}(\omega^{-2}),
\end{equation}
where $t_{1,ij}$ and $t_{2,ij}$ denotes NN and NNN hopping respectively, and
\begin{equation}
t_{1,ij} = \frac{1}{T} \int_0^T {\rm d}\tau \ t_{ij}(\tau), \quad  t_{2, ij} = \sum_{m \neq 0} \sum_k \frac{t_{ik}^{-m} t_{kj}^m}{m\omega}, \quad t_{ij}^m = \frac{1}{T} \int_0^T {\rm d}\tau \ t_{ij} (\tau) e^{-im \omega \tau}.
\end{equation}
The Fourier components of $t_{ij}(\tau)$ can be calculated explicitly. We first rewrite:
\begin{eqnarray}
t_{ij}(\tau) &=& t_{ij} {\rm exp} \big[ -i {\mathcal A}(\tau) \cdot ({\bm r}_j - {\bm r}_i)\big]  \nonumber \\
&=& t_{ij} {\rm exp} \big[ -i A \left( {\rm cos}\omega \tau \ {\rm cos} \phi_{ij} + {\rm sin} \omega \tau \ {\rm sin} \phi_{ij}  \right)\big]   \nonumber   \\
&=& t_{ij} {\rm exp} \big[-i A {\rm cos}(\omega \tau - \phi_{ij}) \big],
\end{eqnarray}
where we have defined the bond angle $\phi_{ij}$ via ${\bm r}_{ij} \equiv {\bm r}_j - {\bm r}_i = ({\rm cos}\phi_{ij}, \ {\rm sin}\phi_{ij})$. Now we can compute the Fourier components:
\begin{eqnarray}
t_{ij}^m &=& \frac{1}{T} \int_0^T {\rm d}\tau \ t_{ij} e^{-iA{\rm cos}(\omega \tau - \phi_{ij})} e^{-im\omega \tau}   \nonumber  \\
&=& \frac{1}{2\pi} \int_0^{2\pi} {\rm d}x \ t_{ij} e^{-i A {\rm sin}x} e^{-imx} e^{-im\phi_{ij}}   \nonumber \\
&=& t_{ij} \mathcal{J}_m(A) e^{-im\phi_{ij}},
\end{eqnarray}
where $\mathcal{J}_m(A)$ is the $m$-th Bessel function of the first kind. It is then straightforward to show $t_{1,ij}= t_{ij} \mathcal{J}_0(A)$ is the renormalized NN hopping, and
\begin{eqnarray}
t_{2,ij}&=& \sum_{m\neq 0} \sum_k \frac{t_{ik}t_{kj}}{m\omega} \ (-1)^m \ \mathcal{J}_m^2(A)\ e^{im(\phi_{ik}-\phi_{kj})} \nonumber  \\
&=& i \sum_k \sum_{m=1}^{\infty}\frac{2t_{ik} t_{kj}}{m \omega} \ (-1)^m \ \mathcal{J}_m^2(A) \ {\rm sin} \left[m(\phi_{ik}-\phi_{kj})\right]  \nonumber \\
&=& -i \eta_{ij} \sum_{m \ {\rm odd}, \ m>0} \frac{4it^2}{m\omega} \ \mathcal{J}^2_m(A) {\rm sin}\left(\frac{\pi}{2}m\right)   \nonumber   \\
&=& \frac{4 \eta_{ij} t^2}{\omega} \sum_{m \ {\rm odd}, \ m>0} \frac{\mathcal{J}_m^2(A)}{m} {\rm sin}\left( \frac{\pi}{2}m\right)
\end{eqnarray}
is the driving-induced NNN hopping, where $\eta_{ij} = \pm 1$ is a sign depending on the hopping direction as well as the even/odd column index of site $i$, as illustrated in Fig. 3(a) in the main text. It is easy to see that the flux through each triangle is $B_{\Delta} = \pm \frac{\pi}{2}$, thus the NNN hopping breaks time-reversal symmetry and gaps out the Dirac cones. We have thus derived the Floquet Hamiltonian (5) in the main text.

\subsection{Edge Theory of $H_{\square}$}

To construct the FT insulator, we couple two copies of $\pi$-flux models with opposite spins under oppositely polarized drives. The FT phases arise when the hoppings satisfy: $t_{1\uparrow} t_{1\downarrow}<0$, $t_{2\uparrow} t_{2\downarrow}<0$. Since the Chern number of the occupied bands in the two copies are $C=\pm 1$, we expect a pair of counter-propagating chiral edge modes at the boundary. However, as we will show in this section, these edge modes can be gapped out by the coupling $H_g$ in the main text. We demonstrate this by constructing the low energy effective edge theory of $H_{\square}$. 

At low energy, we start by linearing $H_{\square}$ around the two Dirac points in momentum space:
\begin{equation}
H  = - v_F ( p_x \sigma^x \otimes \mathbb{1} + p_y \sigma^z \otimes \tau^z ) \otimes \gamma^z + m \sigma^y \otimes \tau^z \otimes \gamma^z,
\end{equation}
where $v_F = 2t_{1}$, $m = 4 t_{2}$, and Pauli matrices $\sigma$, $\tau$ and $\gamma$ act on sublattice, valley and spin degrees of freedom respectively. We shall hereafter omit the $\otimes$. Consider an edge of the system whose normal and tangential unit vectors are ${\bm n} = ({\rm cos}\theta, {\rm sin}\theta)$, ${\bm t} = (-{\rm sin}\theta, {\rm cos}\theta)$. Define ${\bm \Sigma} \equiv (\sigma^x \gamma^z, \sigma^z \tau^z \gamma^z)$, and the effective Hamiltonian near the edge can be written as:
\begin{equation}
H_{\rm edge} = i v_F {\bm n}\cdot {\bm \Sigma} \ \partial_{\lambda} - v_F p_t {\bm t} \cdot {\bm \Sigma} + m(\lambda) \sigma^y \tau^z \gamma^z,
\label{eq:edge}
\end{equation}
where we have decomposed the momentum along the directions of ${\bm n}$ and ${\bm t}$: ${\bm p} = p_n {\bm n} + p_t {\bm t}$, and further made the substitution $p_n \rightarrow -i \partial_{\lambda}$. The edge is modeled by a mass domain wall $m(\lambda)$ interpolating between $\pm m$ as $\lambda \rightarrow \pm \infty$. The edge modes are eigenstates of Hamiltonian~(\ref{eq:edge}) that are exponentially localized near the edge. We take the following ansatz wavefunction for the edge state:
\begin{equation}
\Psi \sim e^{- \frac{1}{v_F} \int_0^\lambda |m(\lambda')| {\rm d} \lambda'} \psi(p_t),
\end{equation}
up to normalization. Here we shall take $m(\lambda') >0$ in the integrand for definiteness. Using this ansatz, we have
\begin{equation}
H_{\rm edge} \Psi = \left[ -v_F p_t {\bm t} \cdot {\bm \Sigma} + m(\lambda) \sigma^y \tau^z \gamma^z (1- i \sigma^y \tau^z \gamma^z {\bm n} \cdot {\bm \Sigma})\right] \Psi.
\end{equation}
If the second term in the above equation vanishes, we have the eigenvalue equation
\begin{equation}
-v_F p_t {\bm t} \cdot {\bm \Sigma} \ \psi(p_t) = E \psi(p_t).
\end{equation}
To make the second term vanish, we simply require that $\psi(p_t)$ is an eigenstate of the projector:
\begin{eqnarray}
P &\equiv& \frac{1}{2} \left( 1 + i \sigma^y \tau^z \gamma^z {\bm n} \cdot {\bm \Sigma}\right)  \nonumber   \\
&=& \frac{1}{2}(1 + \gamma^z {\bm t} \cdot {\bm \Sigma}),
\end{eqnarray}
with $P \psi(p_t) = \psi(p_t)$. Apparently, the projector $P$ and $-v_F p_t {\bm t} \cdot {\bm \Sigma}$ share the common set of eigenstates $\psi(p_t)$. Requiring $\gamma^z {\bm t} \cdot {\bm \Sigma} = 1$ yields the $2 \times 2$ projected edge Hamiltonian:
\begin{equation}
{H}_{\rm edge}^P = - v_F p_t \gamma^z.
\end{equation}
The two counter-propagating chiral edge modes have energies $E = -v_F p_t$ for $\gamma^z=1$ (spin-up), and $E= v_F p_t$ for $\gamma^z=-1$ (spin-down), which are gapless.

Now we demonstrate that the pair of chiral edge modes can be gapped out with $H_g$ in the main text. In momentum space, $H_g$ reads:
\begin{eqnarray}
H_g = &&\sum_{\bm k} g_1 \ c^\dagger_{o{\bm k}, \uparrow} c_{o{\bm k}, \downarrow} - g_1 \ c^\dagger_{e{\bm k},\uparrow} c_{e{\bm k}, \downarrow}  + {\rm H.c.}   \nonumber  \\
&& +\ g_2 \ e^{-ik_x} c^\dagger_{o{\bm k},\uparrow} c_{e{\bm k},\downarrow} - g_2 \ e^{-ik_x} c^\dagger_{o{\bm k}, \downarrow} c_{e{\bm k},\uparrow} + {\rm H.c.},
\end{eqnarray}
where the subscript $e/o$ labels sublattices in even and odd columns respectively. Near the Dirac points, $H_g$ reduces to:
\begin{equation}
H_g = g_1 \sigma^z \gamma^x + g_2 \sigma^x \gamma^y.
\end{equation}
We can now project $H_g$ to the subspace of the edge states, making use of the fact $\gamma^z {\bm t} \cdot{\bm \Sigma} = 1$:
\begin{equation}
P H_g P = g_1 \ {\rm cos} \theta \tau^z \gamma^x -g_2\ {\rm sin}\theta \gamma^y.
\end{equation}
Since both mass terms anticommute with $H_{\rm edge}^P$, we find that indeed they gap out the chiral edge modes. Notice that if only one of the two mass terms is present, either the $x$ or the $y$ edge will remain gapless (i.e. $\theta=0$ or $\theta=\frac{\pi}{2}$), which we also verify numerically.

\section{Appendix D: Fragile Topology of the Floquet $\pi$-flux model}

\subsection{Nontrivial Winding of the Wilson Loop}

\begin{figure}[!b]
(a)
\includegraphics[width=.43\textwidth]{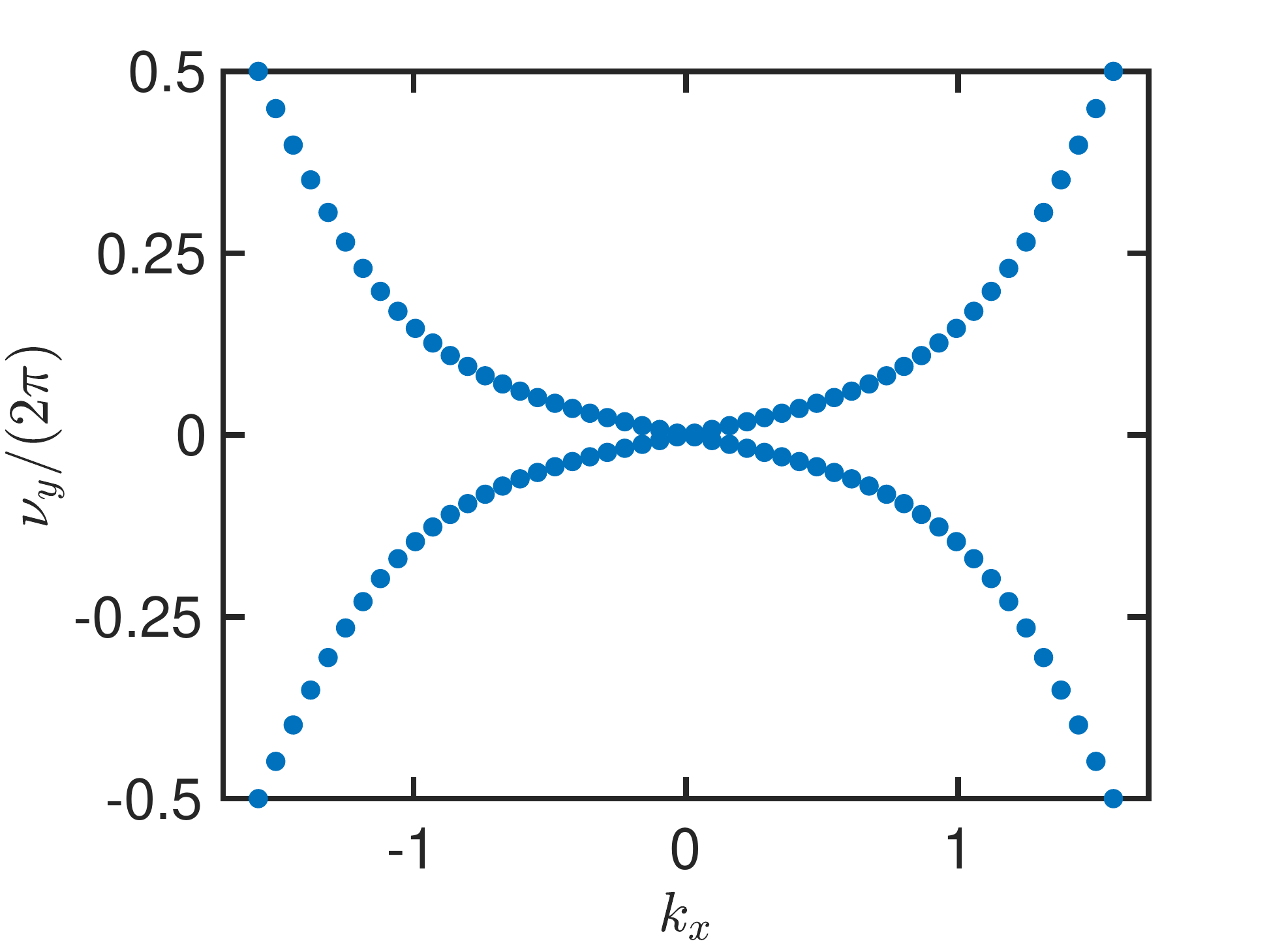}
(b)
\includegraphics[width=.43\textwidth]{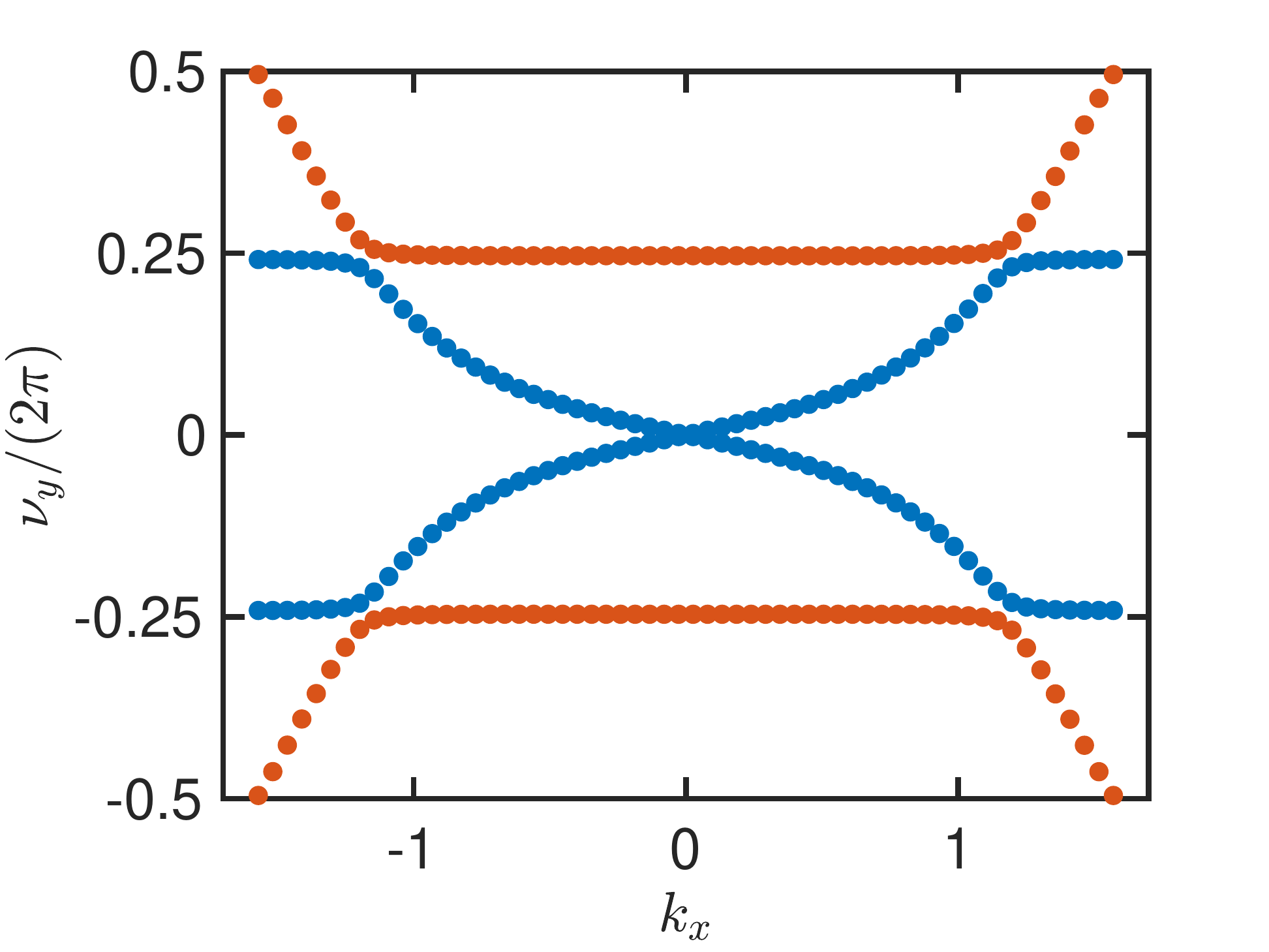}
\caption{(a) Wannier centers of the occupied bands of $H_{\square}$ in phase ${\rm FT}_{\#1}^{\square}$ as a function of $k_x$. The eigenvalues at $k_x = 0$ and $\pm \frac{\pi}{2}$ are protected by the $C_2$ eigenvalues at the high symmetry points of the Brillouin zone. (b) The Wilson loop unwinds upon adding two atomic orbitals at $(0, 1/4)$ and $(0, -1/4)$ to $H_{\square}$. We choose parameters $t_{1\uparrow}=-t_{1\downarrow} = 1$, $t_{2\uparrow}=-t_{2\downarrow} = 0.6$, $g_1=0.3$, $g_2=0.4$, $\mu=10$, $r=0.5$, and $v=1.5$.}
\label{fig:wilson} 
\end{figure}

In this section, we show that the fragile topolgy in the Floquet $\pi$-flux model can be diagnosed from the Wilson loop. 
We take the $FT_{\#1}^{\square}$ phase in Fig. 3(b) of the main text as a concrete example. We compute the Wilson loop oriented along the $k_y$ direction:
\begin{eqnarray}
W_y^{mn} (k_x) &=& \big\langle u^m(k_x, k_y + 2\pi) \big | u^r(k_x, k_y + 2\pi - \Delta k) \big \rangle \cdots \big \langle u^l(k_x, k_y + \Delta k) \big | u^n(k_x, k_y) \big \rangle   \nonumber  \\
&\equiv&  \big\langle u^m(k_x, k_y + 2\pi) \big |  \prod_{\bm k}^{k_y + 2\pi \leftarrow k_y} P({\bm k}) \big | u^n(k_x, k_y) \big \rangle,
\end{eqnarray}
where $P({\bm k})$ is a projector onto the occupied bands. The set of eigenvalues of $W_y$ is denoted as $\{e^{i 2\pi \nu_y(k_x)} \}$, where $\{ \nu_y(k_x) \}$ are the $y$ coordinates of the Wannier centers of the occupied bands at $k_x$. In Fig.~\ref{fig:wilson}(a), we plot the Wannier centers as a function of $k_x$. We find that the Wilson loop exhibits a nontrivial winding across the BZ, indicating an obstruction towards a Wannier representation for the occupied bands of $H_{\square}$ in the ${\rm FT}_{\#1}^{\square}$ phase.

In fact, the winding of the Wilson loop in Fig.~\ref{fig:wilson}(a) is protected by the $C_2$ symmetry of $H_{\square}$. By inspecting the symmetry data in Fig. 3(b) of the main text, we find that the $C_2$ eigenvalues $\Gamma \ (-i,-i)$ and $Y (-i, -i)$ constrain the Wannier center $\nu_y(k_x=0) = 0$, and the eigenvalues $X \ (-i, -i)$ and $M \ (i,i)$ constrain $\nu_y(k_x = \pm \frac{\pi}{2}) = \pm \frac{1}{2}$. Therefore, for $H_{\square}$ with two occupied bands, the Wilson loop winding is robust as long as $C_2$ symmetry is preserved.

\subsection{Addition of Atomic Orbitals}

\begin{figure}[tbh]
(a)
\includegraphics[width=.2\textwidth]{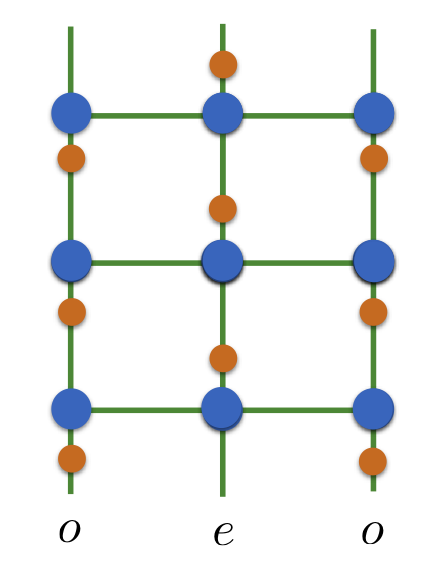}
\quad \quad
(b)
\includegraphics[width=.2\textwidth]{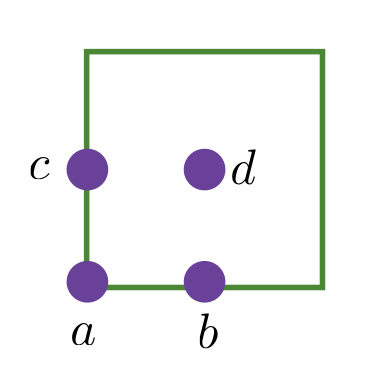}
\caption{(a) Lattice with additional atomic orbitals depicted as orange dots. These orbitals are added in $C_2$ related pairs at positions $(0, \pm 1/4)$ away from the original lattice sites. (b) Four maximal Wyckoff positions of the square lattice.}
\label{fig:atomic} 
\end{figure}

We now show that the nontrivial winding of the Wilson loop can be removed upon adding to $H_{\square}$ atomic orbitals in a $C_2$ symmetric manner. As depicted by orange dots in Fig.~\ref{fig:atomic}, we add atomic orbitals with spin-up electrons at $C_2$ symmetric positions $(0, \pm 1/4)$ away from the original sites. We now couple the additional orbitals to the original model and arrive at the new Hamiltonian:
\begin{equation}
H =
\begin{pmatrix}
H_{\square}  &   H_c \\
H_c^\dagger  & H_{\rm atom}
\end{pmatrix},
\label{eq:trivial}
\end{equation}
where
\begin{equation}
H_{\rm atom} = 
\begin{pmatrix}
-\mu & r e^{-i(k_x+\frac{k_y}{2})}   \\
r e^{i(k_x + \frac{k_y}{2})}   & -\mu
\end{pmatrix},
\end{equation}
and
\begin{equation}
H_c =
\begin{pmatrix}
v e^{-i \frac{k_y}{4}}  &  0 \\
0  & v e^{i \frac{k_y}{4}}  \\
v e^{-i \frac{k_y}{4}}  &  0 \\
0  & -v e^{i \frac{k_y}{4}}  
\end{pmatrix}.
\end{equation}
We have chosen the above form of $H_{\rm atom}$ and $H_c$ such that the $C_2$ symmetry is preserved. 

In Fig.~\ref{fig:wilson}(b), we compute the Wilson loop of the occupied bands of the new Hamiltonian~(\ref{eq:trivial}). We find that the Wilson loop unwinds upon adding trivial atomic orbitals, indicating that the composite system is Wannierizable.

\subsection{Atomic Decomposition}

To further demonstrate that the phases ${\rm FT}_{\#1}^{\square}$ and ${\rm FT}_{\#2}^{\square}$ of $H_{\square}$ are indeed Wannier obstructed, we construct explicitly their decompositions in terms of atomic orbitals. 

\begin{table}[H]
\centering
\begin{tabular}{c|c|c|c|c|c|c|c|c}
\hline
                      &        $s_{\uparrow} @ q_{1a}$       &   $s_{\uparrow} @ q_{1b}$  &   $s_{\uparrow} @ q_{1c}$  & $s_{\uparrow} @ q_{1d}$ &   $s_{\downarrow} @ q_{1a}$ &  $s_{\downarrow} @ q_{1b}$   &   $s_{\downarrow} @ q_{1c}$ &    $s_{\downarrow} @ q_{1d}$                  \\ \hline
  $\Gamma$ &      $-i$          &        $-i$        &     $-i$          & $-i$    &     $i$     &   $i$    &   $i$   &     $i$              \\ \hline
  $X$             &      $-i$           &        $i$         &     $-i$          &  $i$  &  $i$  &  $-i$   &   $i$    &     $-i$                 \\ \hline
  $Y$             &       $-i$          &         $-i$      &      $i$          &   $i$   & $i$   &  $i$     &    $-i$   &    $-i$           \\ \hline
  $M$           &        $-i$          &      $i$          &      $i$           &  $-i$   & $i$   &  $-i$   &   $-i$   &    $i$                  \\ \hline
\end{tabular}
\caption{$C_2$ eigenvalues at high symmetry momenta obtained by putting atomic orbitals at four maximal Wyckoff positions. We only consider $s$ orbitals with two spin species.}
\label{table:atomic}
\end{table}

In Table~\ref{table:atomic}, we list the $C_2$ eigenvalues at high symmetry momenta obtained from putting atomic orbitals at four maximal Wyckoff positions of the square lattice [Fig.~\ref{fig:atomic}(b)]. We consider only $s$ orbitals with two spin species. Comparing with Fig. 3(b) in the main text, we find that the fragile topological phases of $H_{\square}$ can be decomposed as follows:
\begin{eqnarray}
{\rm FT}_{\#1}^{\square} &=& (s_{\uparrow}@q_{1a}) \oplus (s_{\uparrow} @ q_{1b}) \oplus (s_{\uparrow} @ q_{1c})  \ominus (s_{\uparrow} @ q_{1d}),    \\
{\rm FT}_{\#2}^{\square} &=& (s_{\downarrow} @ q_{1a}) \oplus (s_{\downarrow} @ q_{1b}) \oplus (s_{\downarrow} @ q_{1d}) \ominus (s_{\downarrow} @ q_{1c}).
\end{eqnarray}
Indeed, both ${\rm FT}_{\#1}^{\square}$ and ${\rm FT}_{\#2}^{\square}$ can be represented as subtracting an atomic insulator from another atomic insulator, which indicates that both phases exhibit fragile topology.

Upon adding atomic orbitals and forming the composite Hamiltonian~(\ref{eq:trivial}), the system becomes Wannierizable and hence should be representable in terms of atomic orbitals. We take ${\rm FT}_{\#1}^{\square}$ as an example. 
The symmetry data for Hamiltonian~(\ref{eq:trivial}) by coupling with ${\rm FT}_{\#1}^{\square}$ are: $\Gamma \ (-i, -i, i, -i)$, $X \ (-i, -i, i, -i)$, $Y \ (-i, -i, i, -i)$, and $M \ (i, i, i, -i)$. By inspecting Table~\ref{table:atomic}, we find that the composite system can be represented as $(s_\uparrow @ q_{1a}) \oplus (s_\uparrow @ q_{1b}) \oplus (s_\uparrow @ q_{1c}) \oplus (s_\downarrow @ q_{1d})$, which is indeed an atomic insulator.

\subsection{Mirror-protected Fragile Topology of the Floquet $\pi$-flux Model}
Along the mirror-symmetric line in Fig. 3(c) of the main text, the system further preserves mirror symmetries $M_x$ and $M_y$. In this case, atomic decompositions must also take into account mirror symmetry representations, which leads to new mirror-protected fragile phases, mirror-FT$_{\#1}^\square$ and mirror-FT$_{\#2}^\square$. Since $[M_x,M_y]=0$, the symmetry data consist of ($m_x,m_y$) at each high symmetry momenta, which is a pair of simultaneous eigenvalues of $M_x$ and $M_y$. We list the symmetry data for our target fragile phases and possible atomic insulators in Table. \ref{table:mirror}. For the atomic phases in Table. \ref{table:mirror}, we have only listed cases with an atomic orbital carrying $m_x=+i$, while the situation with $m_x=-i$ can be derived similarly by simply flipping the sign of $m_x$.


\begin{table}[t]
	\includegraphics[width=\textwidth]{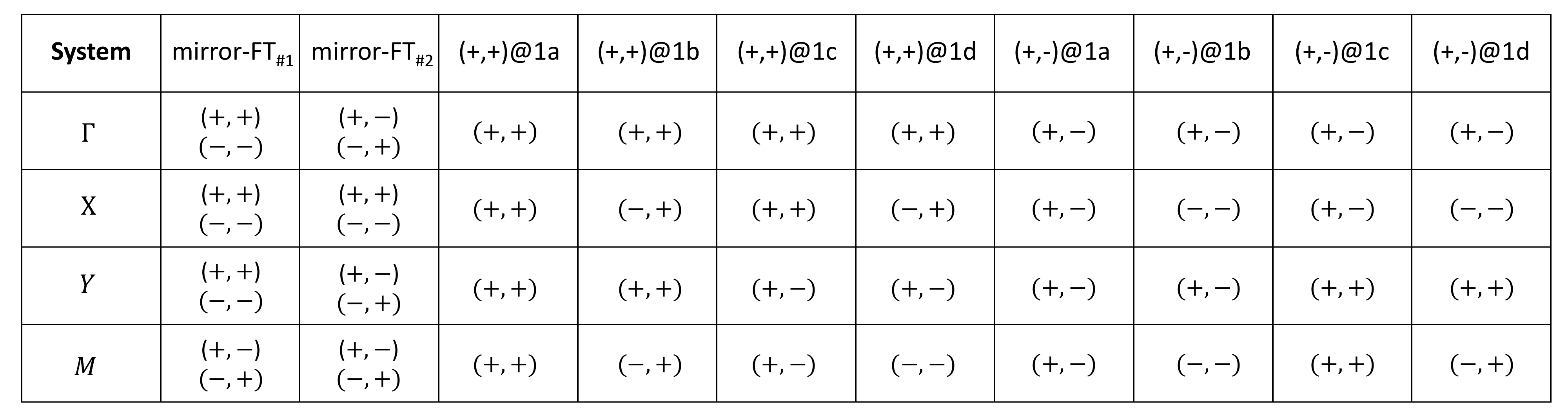}
	\caption{Mirror eigenvalues $(m_x,m_y)$ at high symmetry momenta for mirror-${\rm FT}_{\#1}$ and mirror-${\rm FT}_{\#2}$ and various atomic insulators. Here we denote $(\pm i,\pm i)=(\pm,\pm)$ for short.}
	\label{table:mirror}
\end{table}

By comparing with the atomic data, we again find that no two-band atomic insulator could match the symmetry data for either mirror-FT$_{\#1}^\square$ or mirror-FT$_{\#2}^\square$, which implies the existence of Wannier obstruction. On the other hand, we find that the two phases yield the following atomic decompositions:
\begin{eqnarray}
\text{mirror-FT}_{\#1} &\equiv& (-,-)@q_{1a} \oplus (+,+)@q_{1b} \oplus (+,+)@q_{1c} \ominus (+,+)@q_{1d}, \nonumber \\
\text{mirror-FT}_{\#2} &\equiv& (-,+)@q_{1a} \oplus (+,-)@q_{1b} \oplus (+,+)@q_{1c} \ominus (+,+)@q_{1d},
\end{eqnarray} 
which indicates mirror-protected fragile topology.

\end{document}